\documentclass[12pt]{article}

\usepackage{scicite}

\usepackage{amsmath}
\usepackage{times}
\usepackage{txfonts}
\usepackage[]{graphicx}
\usepackage{booktabs}
\usepackage{subcaption}
\usepackage{xcolor}
\usepackage{lineno}
\usepackage{comment}
\usepackage{subcaption}
\usepackage{hyperref}
\usepackage[normalem]{ulem}
 \def\mso{\,\mathrm{M_\odot}}

 \def\lso{\,\mathrm{L_\odot}}
 \def\Msun{\,\mathrm{M_\odot}}

 \def\kms{\, \mathrm{km\,s^{-1}}}

 \def\Yi{Y_\mathrm{i}}
 \def\Ys{Y_\mathrm{s}}
 \def\Mi{M_\mathrm{i}}
 \def\Me{M_\mathrm{e}}
 \def\Menv{M_\mathrm{env}}
 \def\Ycno{Y_\mathrm{CNO}}
 
 \def\Macccno{M_\mathrm{acc,CNO}}

 \def\fcno{f_\mathrm{CNO}}
 \def\fcn{f_\mathrm{CN}}
 
 \def\logTeff{\log \Teff \, \mathrm{[K]}}
 \def\logLLsun{\log L / L_\odot}
 \def\Pi{P_\mathrm{i}}
 \def\qi{q_\mathrm{i}}
 \def\Mpi{M_\mathrm{1,i}}
 \def\Msi{M_\mathrm{2,i}}
 \def\Teff{T_\mathrm{eff}}
 \def\logg{\log g}
   
 \def\vsini{v\,\mathrm{sin}\,i}

 \def\NO{\left(\frac{N}{O}\right)}
 
 \def\CO{\left(\frac{C}{O}\right)}
 \def\OC{\left(\frac{O}{C}\right)}

 \def\Ci{\mathrm{C_i}}
 \def\Ni{\mathrm{N_i}}
 \def\Oi{\mathrm{O_i}}
 \def\Ce{\mathrm{C_e}}
 \def\Ne{\mathrm{N_e}}
 \def\Oe{\mathrm{O_e}}

 \def\Menv{M_\mathrm{env}}

 \def\logNO{\log \mathrm{(N/O)}}

\newcommand{\Fig}[1]{Fig.~\ref{#1}}

  \def\simle{\mathrel{\hbox{\rlap{\hbox{\lower4pt\hbox{$\sim$}}}\hbox{$<$}}}}
 \def\simgr{\mathrel{\hbox{\rlap{\hbox{\lower4pt\hbox{$\sim$}}}\hbox{$>$}}}}

\newcommand*\xbar[1]{%
   \hbox{%
     \vbox{%
       \hrule height 0.5pt 
       \kern0.5ex%
       \hbox{%
         \kern-0.1em%
         \ensuremath{#1}%
         \kern-0.1em%
       }%
     }%
   }%
}

\newcommand*{\rev}[1]{\textcolor{black} {#1}}
\newcommand*{\edi}[1]{\textcolor{black} {#1}}

\usepackage{ulem}
\usepackage{color}
    \newcommand{\araa}{Annu. Rev. Astron. Astrophys.}   
    \newcommand{\aj}{Astron. J.}   
    \newcommand{\apj}{Astrophys. J.}   
    \newcommand{\apjl}{Astrophys. J. Lett.}   
    \newcommand{\apjs}{Astrophys. J. Suppl. Ser.}   
    \newcommand{\aap}{Astron. Astrophys.}   
    \newcommand{\mnras}{Mon. Not. R. Astron. Soc.}   
    \newcommand{\nat}{Nature} 
    \newcommand{\nastro}{Nat. Astron.} 
    \newcommand{\pasp}{Publ. Astron. Soc. Pac.}   

\newenvironment{sciabstract}{%
\begin{quote} \bf}
{\end{quote}}

\title{Chemical fingerprints of binary mass transfer in massive stars}
\author{Harim Jin$^{1,2}$, Norbert Langer$^{2,3}$\\
\\
\footnotesize{$^{1}$Max Planck Institute for Astrophysics, Karl-Schwarzschild-Straße 1, 85748 Garching bei München, Germany}\\
\footnotesize{$^{2}$Argelander-Institut f\"ur Astronomie, Universit\"at Bonn, Auf dem H\"ugel 71, 53121 Bonn, Germany}\\
\footnotesize{$^{3}$ Max-Planck-Institut f\"ur Radioastronomie, Auf dem H\"ugel 69, 53121 Bonn, Germany} \\
\\
\footnotesize{$^\ast$To whom correspondence should be addressed; E-mail: jin@mpa-garching.mpg.de}}
\date{\today}

\newcommand{\bibcommenthead}{}
\begin{document} 

\baselineskip24pt

\maketitle 
\bigskip

\begin{sciabstract}

{\small The majority of massive stars are born in close binary systems. As stars expand when they age, mass transfer or even a merger with their companion is inevitable. However, most binary interaction products appear as single stars, such that the main evidence of their exciting past is lost. In a comprehensive grid of detailed massive binary evolution models we find systematic trends in chemical surface abundances that allow identifying the past mass gainers. We develop an analytic framework which is independent of specific evolutionary models, to constrain the amount and composition of the accreted material from their observed surface abundances. This yields tight constraints on the uncertain mass transfer physics in massive binary stars and allows us to reconstruct the past evolutionary history of the progenitor binary system. This method, which is shown to also constrain binary mergers (for example, SN 1987A), is applied to some of the best-studied OB stars so far. For $\gamma\,$Columbae, suggested to be an envelope-stripped star, we show that it is a mass gainer instead, whose companion star likely formed a stripped-envelope supernova. Our results highlight surface abundance measurements as a powerful tool to improve our understanding of massive binary systems evolving towards supernovae and compact object binaries.}

\end{sciabstract}

The fraction of unevolved massive stars with a stellar companion so close that mass exchange becomes inevitable is $\sim$70\%, independent of the heavy element fraction of the star forming environment \cite{Sana2025}. With less than \rev{$\sim$0.1\%} of their lifetime, the main episode of mass exchange is typically so fast that it is very rarely caught by observations. At the same time, it takes far too long for hydrodynamic models to assess it entirely (cf. \cite{Ryu2025}). Therefore, the stability of the mass transfer process, as well as its efficiency (the fraction of mass that will be accreted by the companion star), 
remain weakly constrained so far, e.g. through population models of post-interaction stars \cite{Vinciguerra2020,Xu2025,Schuermann2025}. Binary mass transfer changes the evolution and fate of the involved stars drastically, even often leading to a merger of both components, or changing whether and how the stars explode, and what final remnants they produce. This affects our understanding of the chemical, mechanical, and radiative feedback of massive stars \cite{Goetberg2020,Farmer2023,Wagg2025}, which shapes the evolution of star-forming galaxies across cosmic time. 

Empirical constraints for the mass transfer process have been worked out for individual cases where the binary nature of the post-mass transfer system could be identified \cite{Shenar2016, Sen2022, Lechien2025}. However, this concerns small samples, which do not allow a comprehensive assessment of the dependence of the mass transfer physics on the binary parameters. Observationally, the vast majority of mass gainers and stellar merger products are not recognized as such --- they mostly appear single because they have lost their companion, merged with it, or the companion is faint and of low mass \cite{deMink2014,Goetberg2017}. Up to $\sim$30\% of all core hydrogen burning and $\sim$70\% of core helium burning stars are expected to have a binary interaction history \cite{Sana2012,deMink2014}. At the same time, a large fraction of massive stars are found to show thermonuclear processed matter at their surface, which provides a large reservoir of information.
However, so far it has proven difficult to distinguish between the different possible sources of enrichment, which are stellar wind mass loss, envelope mixing in single stars, and binary interaction \cite{Meynet2000,Hunter2008,Langer2012,Brinkman2025}.

\textbf{Results from detailed binary evolution models}

\begin{figure*}
	\centering
	\includegraphics[width=0.75\linewidth]{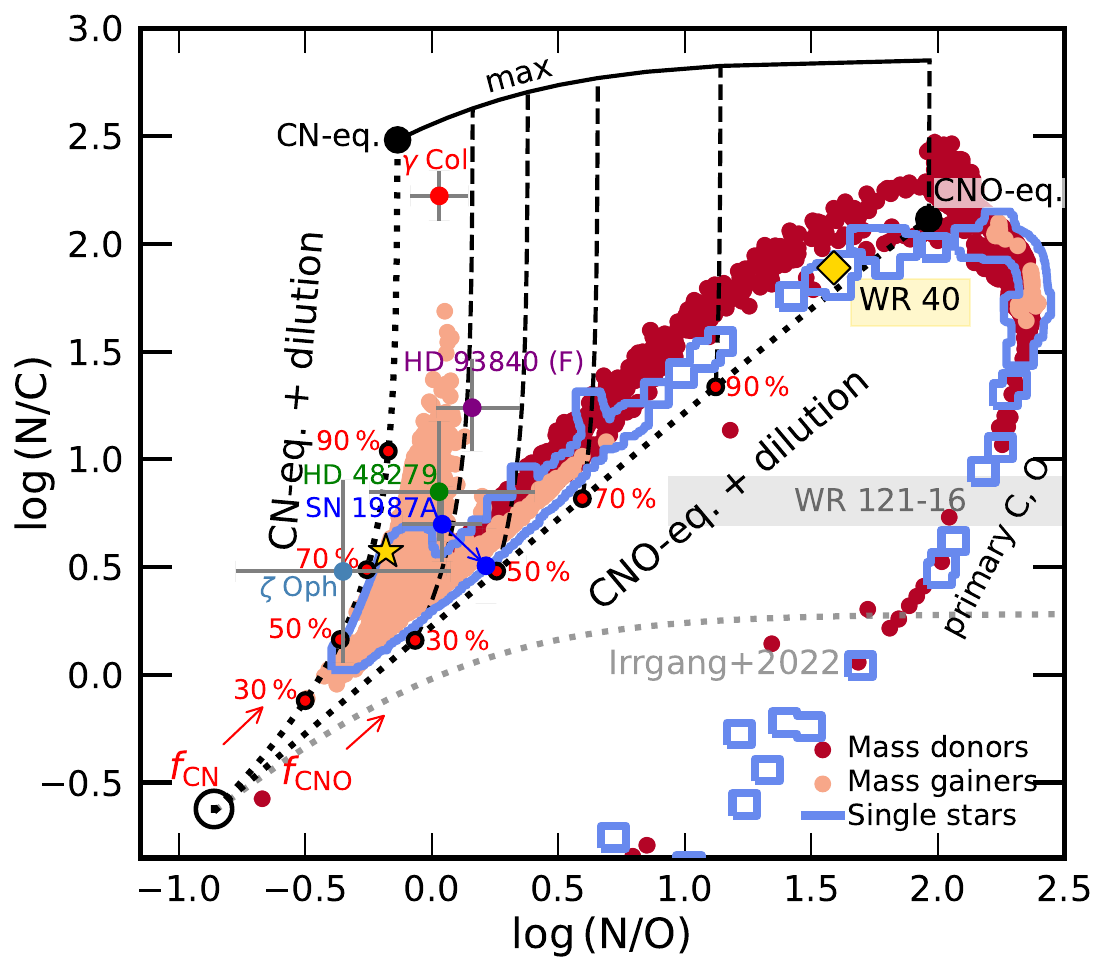}
    \caption{{\bf Diagnostic CNO-surface abundance diagram.} N/C and N/O surface abundance ratios of mass gainer (light red) and mass donor (dark red) models after mass transfer, and single star models of \cite{Jin2024} with initial masses between 5 and 50$\mso$ and initial rotational velocities below 400$\kms$ (blue contours) during core helium burning. The two dotted black lines represent analytic expressions for the composition of a mixture of pristine matter with matter containing CN- (Eq.\,\ref{eq_cn}), or CNO-equilibrium abundances (Eq.\,\ref{eq_cno}), respectively \cite{Langer2012}. Their common bottom left edge, marked by a black circle, reflects the initial composition of our models, and their upper right ends marked by black filled circles correspond to the pure equilibrium composition. Several values of the dilution factors, $\fcn$ and $\fcno$, are marked by red circles along the dotted lines. The dashed lines equidistant to the ``CN-eq. + dilution''-line are obtained by assuming CN-processing, starting with a composition at the ``CNO-eq. + dilution''-line (Eq.\,\ref{eq_fcno_cn}). They end at the top black full-drawn line, which represents complete CN-processing (Eq.\,\ref{eq_max}). \edi{The colored filled circles with error bars} represent observed values for stars HD\,48279 \cite{Martins2015}, $\gamma\,$Columbae \cite{Irrgang2022}, HD\,93840 \cite{Wessmayer2024}, $\zeta$ Ophiuchi \cite{Villamariz2005}, and the peculiar Type IIP supernova SN\,1987A \cite{Lundqvist1996} (see Supplementary Information Section A), \edi{where error bars denote 1$\sigma$ uncertainties, except for $\gamma$ Columbae (99\% confidence interval).} The arrow pointing away from the symbol for SN\,1987A reflects the required shift to correct for the non-solar CNO abundance ratios in the Large Magellanic Cloud (Supplementary \Fig{fig_LMC}). The yellow star symbol represents our ``Mock star'' (see text). The yellow diamond marks the WN8 star WR\,40 \cite{Herald2001}, and the grey rectangle the WN/C star WR\,121-16 \cite{Zhang2020}. The grey dotted line replicates the line designated as ``ON-cycle limit'' in fig.\,2 of \cite{Irrgang2022}, which is based on the analytical expression of eq.\,(14) in \cite{Maeder2014}.}
	\label{fig_CNO2}
\end{figure*}

It has recently become possible to compute comprehensive grids of massive binary evolution models, which densely cover the whole initial binary parameter space, and include a detailed treatment of internal mixing and nucleosynthesis \cite{Langer2020}. Here, we employ such a grid computed with mass and angular momentum transfer, differential rotation, and tides \rev{using MESA \cite{Paxton2011,Paxton2013,Paxton2015,Paxton2018,Paxton2019}}. The models were computed with an extended nuclear network through hydrogen burning which follows the time evolution of all stable CNO isotopes \cite{Jin2025}, which is necessary to accurately compute deviations from CN- and CNO-equilibrium \edi{(see Extended Data Table~\ref{tab_cno})}. We scan through the parameter space of the mass gainer models in search for distinctive surface abundance patterns of binary interaction products. We focus on the elemental abundances altered by the CNO-cycle, i.e. helium, carbon, nitrogen, and oxygen. Heavier elements (e.g. sodium) and isotopic ratios are also affected but are harder to observe. We focus on the mass gainers that experience accretion during their main sequence stage, become the brighter of the two stars, and still live for a long time. The binary model grid does not include mergers, but we discuss their expected surface abundances based on simplified models below. 

Figure\,\ref{fig_CNO2} shows the CNO surface abundances of the mass gainer models during core helium burning, which reflects their enrichment at core hydrogen exhaustion (cf. Extended Data Fig.~\ref{fig_CNO_GENEVA} for the evolution through the core hydrogen burning stage), together with those of mass donors and single stars. Since the CN- and CNO-cycles conserve the number of CNO nuclei, our models cannot populate the region outside of the area framed by the two equilibrium lines (black dotted lines), unless primary CNO products caused by $\alpha$-captures appear at the surface when helium burning products are revealed. Surprisingly, Fig.~\ref{fig_CNO2} also shows that most of the allowed area remains empty, as the models populate only two branches, one close to the ``CNO-eq. + dilution''-line and the other close to the ``CN-eq. + dilution''-line. 

Finding models on the lower of the two branches is expected, since envelope stripping, by winds or by mass transfer, reveals matter in the former convective core of the stripped stars, for which the high central temperature of the core ensures CNO-equilibrium. This branch is therefore populated by mass donors of any initial mass, and by the mass gainers and single stars that are so massive that they can self-strip their envelopes by winds and evolve to become hydrogen-poor, so called Wolf-Rayet stars.

Strikingly, the branch near the ``CN-eq. + dilution''-line is exclusively populated by mass gainers and is avoided by mass donors and single stars. This is surprising at first because, next to pristine matter from the donor's envelopes, the mass gainers accrete matter containing CNO-equilibrium abundances from the mass donors. 
However, this can be understood as follows. At low enough temperatures, CN-equilibrium may be reached while CNO-equilibrium is not \cite{Rolfs1988}, as in the core of our Sun ($\sim$15\,MK). In massive core-hydrogen burning stars, CNO-equilibrium is quickly obtained in their cores ($\sim$30$\dots$40\,MK). In addition, in a small region just above the convective core, temperatures resemble those in the Solar center ($\sim$15$\dots$17\,MK) and allow for the conversion of carbon to nitrogen via the CN cycle, but CNO-equilibrium can not be achieved. In single stars, this usually remains without observable consequences. 

In our mass gainers, matter in CNO-equilibrium with and without helium enrichment is added to the envelope, besides pristine matter. The most enriched matter is added last, and the surface helium- and nitrogen enrichment is very high at this time (Extended Data Fig.\,\ref{fig_detail}). The helium enrichment of the outer layers imposes fast thermohaline mixing which evenly distributes all chemical components in the mass gainer's envelope on the thermal timescale (``fast mixing''; $\tau_{\rm th}\simeq 10^4\,$yr). The timescale of this fast mixing is much shorter than the timescale of CN-processing at the bottom of the envelope ($\tau_{\rm CN}\simeq 0.5\,$Myr), such that the envelope is not enriched in CN-processed matter at this stage. The star will therefore be close to the ``CNO-eq. + dilution''-line in Fig.\,\ref{fig_CNO2} at this time (see also Extended Data Fig.\,\ref{fig_mock}).  After accretion and fast envelope mixing, thermohaline mixing persists during the further nuclear timescale evolution, but as the inverted chemical composition profile is nearly smoothed out, the mixing timescale eventually exceeds the timescale of CN-processing at the bottom of the envelope (``slow mixing''). Assisted by rotational mixing in the spun-up envelope, this slow mixing brings unprocessed carbon into the CN-cycling layer at the bottom of the envelope,
and CN-processed matter towards the surface, where the N/C-ratio increases. If CN-processing and mixing were efficient enough, the star could reach the top black line in Fig.\,\ref{fig_CNO2}, but not higher N/C-values.

The described slow mixing elevates many of our mass gainer models in Fig.~\ref{fig_CNO2} significantly above the ``CNO-eq. + dilution''-line, while their envelope mixing remains incomplete. Their evolution in Fig.~\ref{fig_CNO2} during this phase can be well approximated by the dashed black lines. They start on the ``CNO-eq. + dilution''-line (Eq.~\ref{eq_cno}) at a point which reflects the amount of accreted CNO-equilibrium matter, and then follow the analytic prescription for converting the available carbon to nitrogen (Eq.~\ref{eq_fcno_cn}) towards CN-equilibrium (Eq.~\ref{eq_max}, see Methods). For an observed star located above the ``CNO-eq.+dilution'' line, the black dashed lines can trace its evolution back to its foot point on the ``CNO-eq.+dilution'' line, to determine the accreted fraction of CNO-equilibrium material. 

Notably, the efficiency of thermohaline mixing in massive stars has so far not been calibrated empirically\rev{, nor has its interplay with rotational mixing been fully understood}, such that the elevation of our gainer models above the ``CNO-eq. + dilution''-line must be considered uncertain (cf., Extended Data Fig.\,\ref{fig_mock}). 
While the described mechanism has been discussed before \cite{Braun1997, Marchant2017, Renzo2021}, it has so far not been evaluated in a population of binary models, nor was its unique CNO-signature worked out. Figure\,\ref{fig_CNO2} indicates that the CNO signature can be used as a fingerprint to identify past mass gainers observationally. Below we show that it allows to quantitatively constraint their past accretion history, independent of uncertainties in the efficiency of the slow mixing in the envelope of stellar models.

\textbf{Analytic framework}
\begin{figure}
	\centering
	\includegraphics[width=0.9\linewidth]{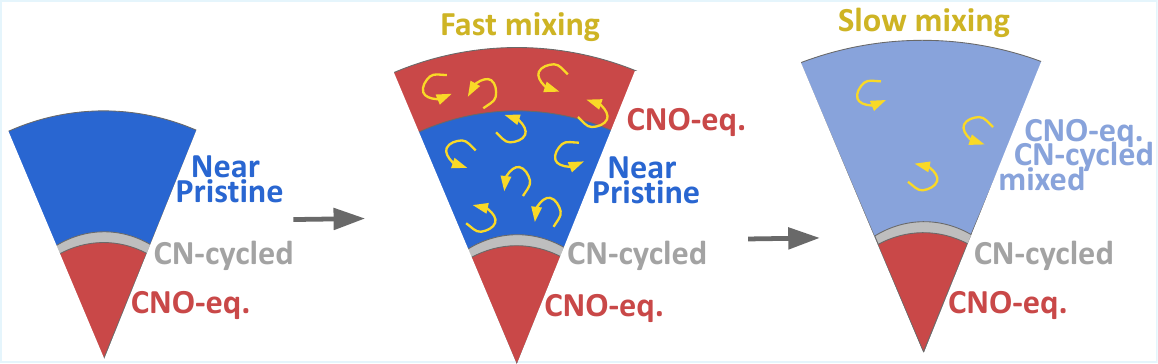}
	\caption{{\bf Schematic illustration of the chemical evolution of a mass gainer.} Three stages are shown: before mass accretion, immediately after mass accretion just before fast mixing, and during slow mixing in the subsequent nuclear-timescale evolution. The radial direction of each slice represents the encompassed mass.}
	\label{fig_pie}
\end{figure}

\begin{figure}
	\centering
	\includegraphics[width=0.8\linewidth]{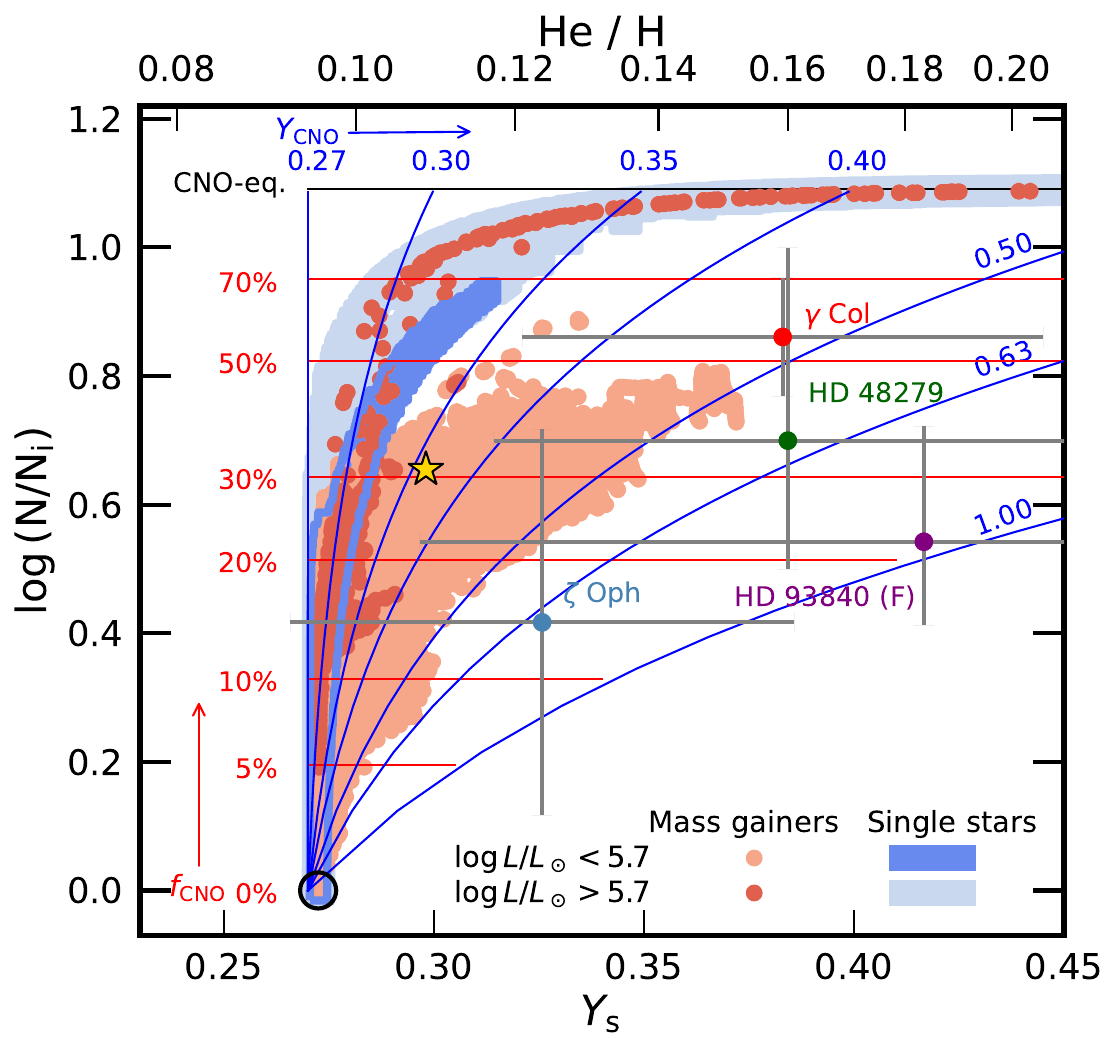}
	\caption{{\bf Diagnostic helium-nitrogen surface abundance diagram.} Helium mass fraction and nitrogen enhancement factor at the surface of our mass gainer models after the accretion event, until core hydrogen exhaustion (filled light and dark red circles, issued every 100\,000 years). Also shown are rotating and mass losing single star models from \cite{Jin2024} with initial rotational velocities below 400$\kms$ (light and dark blue). We distinguish models with luminosities of $\logLLsun < 5.7$, which do not uncover their initial convective core due to strong wind mass loss during the main sequence evolution. The black horizontal line at the top marks the nitrogen enhancement factors corresponding to CNO-equilibrium, and red horizontal lines indicate by which factor $\fcno$, given at the left side, CNO-equilibrium matter is diluted with pristine matter to achieve the given enhancement level without CN-cycling (Eq.\,\ref{eq_YNfcno}). The blue lines represent abundances assuming mixing of pristine matter and CNO-equilibrium matter with a helium mass fraction of $\Ycno$ (Eq.\,\ref{eq_NY}), given at the right side or the top of the line. Their common bottom left edge, marked by a black circle, reflects the initial composition of our models. The upper $x$-axis shows the surface helium-to-hydrogen number ratio. \edi{The colored filled circles with error bars} represent HD\,48279 \cite{Martins2015}, $\zeta$ Ophiuchi \cite{Villamariz2005}, $\gamma$ Columbae  \cite{Irrgang2022}, and HD\,93840 \cite{Wessmayer2024}, \edi{where error bars denote 1$\sigma$ uncertainties, except for $\gamma$ Columbae (99\% confidence interval).} The star symbol represents our ``Mock star'' (see text). Most of the depicted mass gainers models (93\%) are either single or have a degenerate companion, and would observationally appear as single stars.}
	\label{fig_HeN}
\end{figure}

We set up a simple, analytic model to exploit the quantitative information contained in the CNO diagram. 
We consider mass transfer to occur when the initially more massive star (primary) attempts to expand right after core hydrogen exhaustion (Case\,B). At this stage, the companion (secondary) is burning hydrogen in its core, and the interiors of both stars can be divided into a helium-rich core, a near pristine hydrogen-rich envelope (with a small CN-processing region at the bottom), and in-between a layer containing the hydrogen/helium gradient left by the receding convective core in the course of hydrogen burning. Notably, the composition in the inner two regions is in CNO-equilibrium in both stars.  

We then assume that the mass gainer accretes parts of the pristine envelope and the H/He-gradient layer of the mass donor, which is instantly mixed with the gainer's envelope \edi{(cf. Extended Data Figs.~\ref{fig_schematic} and~\ref{fig_Yderivation})}. This alone would place the mass gainers on the ``CNO-eq. + dilution''-line in Fig.~\ref{fig_CNO2}, with their location set by the fraction of the H/He gradient layer incorporated into the envelope ($\fcno$). In the final step, we assume slow mixing of the envelope and an increase of the N/C surface ratio due to conversion of the available carbon to nitrogen by the CN-cycle \edi{(\Fig{fig_pie})}, along a line parallel to the dashed lines in Fig.~\ref{fig_CNO2}. The efficiency of this last process determines the final N/C-ratio. As we explain in Methods, all steps can be worked out analytically. 

Before applying this model to an observed star, it is interesting to consider the location of the stellar models in the helium-nitrogen (HeN) diagram to unlock the potential of helium surface abundance measurements \cite{Carlos2025}. Figure~\ref{fig_HeN} shows that, also here, the mass gainers are found in a distinct sub-space of the diagram, although rotating single star models in which core material is mixed efficiently into the envelope may populate a similar area (see Extended Data Fig.~\ref{fig_CNO_GENEVA}). The strength of this diagram is that it allows to determine the average helium abundance in the fraction of the H/He-gradient layer of the donor that is accreted by the mass gainer ($Y_{\rm CNO}$). 

However, since surface nitrogen is enhanced not only by the accretion of CNO-equilibrium matter, but also by  subsequent CN-cycling and slow mixing in the mass gainer's envelope, only upper limits on $\fcno$ can be derived from this diagram alone (cf. Extended Data \Fig{fig_mock}).
Only in combination with the CNO diagram, the exact amounts can be determined, and the ambiguity with single star models is lifted. 

We emphasize that our simple model is motivated by the findings from our numerical binary evolution models, but the inferred amounts of accreted CNO-equilibrium material and helium (see Methods) are determined independently of those models. 
\rev{The physics which sets the accretion efficiency in our numerical MESA models had to be fixed before the binary model grid was computed.} Accordingly, the accretion efficiency derived for an obtained progenitor configuration of observed \rev{accretors via our analytic framework} generally differs from the accretion efficiency of the corresponding binary evolution models in the binary model grid. \rev{These differences will be useful as empirical benchmarks for future numerical binary evolution models.}

With three measured quantities, we cannot directly invert the equations, as we have five unknowns (the two initial masses, the masses of pristine and CNO-cycled accreted matter, and the average helium abundance in the latter). However, since different initial masses produce different amounts of CNO-processed material in the H/He gradient layer, this provides additional constraints on the initial mass of the donor star. Similarly, the current mass of the gainer, combined with the amount of accreted CNO-equilibrium material, constrains the initial mass of the mass gainer. Further constraints arise from considering the timing of the donor's supernova explosion, and the requirement of stable mass transfer, which depends on the initial mass ratios of the binary (see Methods).

We test our simple model by applying it to one of our post-mass transfer MESA models that initially consisted of $22.4\mso + 7.8\mso$ stars with an initial orbital period of 10 days. This model leads to an early B-type Mock star which is characterized by relatively high surface abundance ratios, with N/O$\approx$1 and N/C$\approx$4, after Case B mass transfer (see Extended Data Table~\ref{tab_obs} for a list of parameters and Extended Data \Fig{fig_mock} for its evolutionary history in the abundance planes). Our method successfully reproduces the key properties of the accreted material such as $\Macccno$, and the initial binary configuration.

\textbf{Application to observed stars}

\begin{figure*}[h!]
	\centering
	\includegraphics[width=0.70\linewidth]{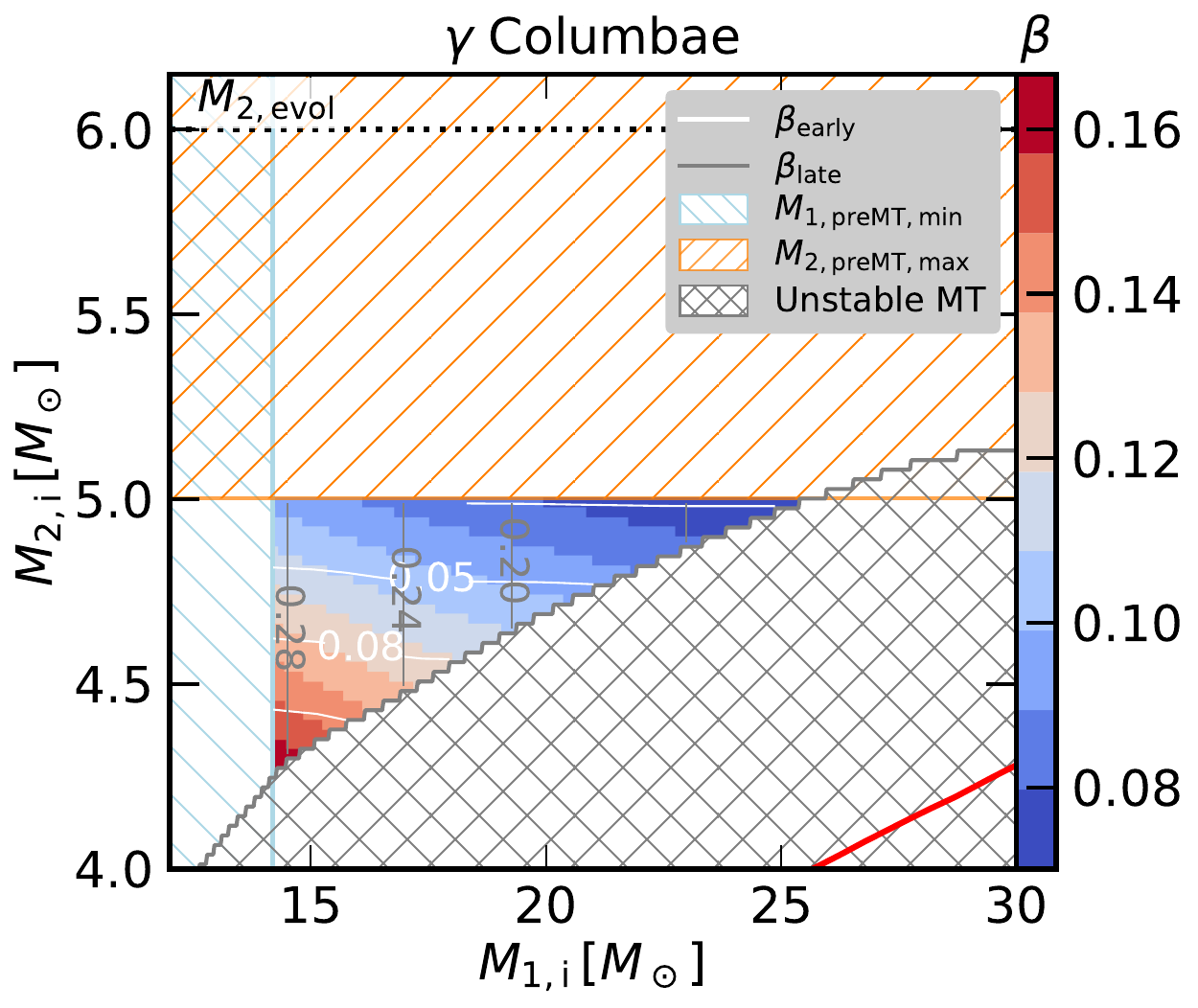}
	\caption{{\bf Initial masses diagram for $\gamma\,$Columbae.} In the parameter space of the initial masses ($\Mpi$ and $\Msi$) of the components of the binary system that produced $\gamma\,$Col, we overplot the various constraints indicated by different hatchings (see Methods). The colored nearly triangular region contains the remaining possible initial configurations, where the color represents the average mass accretion efficiency \edi{($\beta$)} during the mass transfer (see color bar to the right). White and gray contour lines in this region denote allowed accretion efficiencies for the transfer of pristine ($\beta_{\rm early}$) and CNO-processed ($\beta_{\rm late}$) matter, respectively. The black horizontal dotted line indicates the current mass of $\gamma\,$Col. The thick red line marks where $\beta_\mathrm{early}$ and $\beta_\mathrm{late}$ are the same.}
    \label{fig_gammacol}
\end{figure*}

To demonstrate the benefits of our method, we use it for some of the best studied OB stars, which were found to exhibit significant surface abundance enrichment. A well observed case is the naked eye star $\gamma$\,Columbae of spectral type B2.5IV, an apparent single star of $\sim$6.0$\mso$. In a recent spectral analysis, a helium surface mass fraction of $0.38\pm0.06$ and a nitrogen enhancement factor of $\sim$7 were found \cite{Irrgang2022}. Interpreting it as a former binary mass gainer (see Supplementary Information Section B), we find that, based on its CNO-fingerprint, its envelope material displays nearly undiluted CN-equilibrium abundances (Fig.~\ref{fig_CNO2}). With a current envelope mass of $4.4\mso$ (Extended Data Table~\ref{tab_obs}), its N/O-value indicates that about $0.8\mso$ ($\fcno\sim$17\%) of matter containing CNO-equilibrium matter had been accreted. Shifting the observational point for $\gamma\,$Columbae down to the corresponding $f_{\rm CNO}$-value of 17\% reveals that the average helium mass fraction in this matter was $\sim$90\% (Fig.~\ref{fig_HeN}).

The initial parameters of the progenitor binary of $\gamma\,$Columbae can now be constrained as follows. 
With an evolutionary mass of 6.0$\mso$ and having accreted at least 0.8$\mso$, the initial mass of $\gamma\,$Columbae must have been smaller than $5.2\mso$ (see Methods for an additional constraint giving $<5.0\mso$). The mass donor must have contained a region of 0.8$\mso$ in its H/He-gradient region with an average helium mass fraction of 90\% prior to mass transfer. Investigating single star models at core hydrogen exhaustion reveals that this excludes initial donor masses below $\sim$ 14$\mso$ (see Methods for more details). This implies that the initial mass ratio of the binary was below 0.35, and that the mass transfer was highly non-conservative.

The initial binary parameter space for $\gamma\,$Columbae can be further constrained since at a given mass accretion rate, mass transfer becomes unstable for too small initial accretor masses, as the accretor would swell and the binary would merge. This process has recently been extensively studied \cite{Schuermann2024}. Using their results allows us to check for each point in the initial masses-diagram (Fig.~\ref{fig_gammacol}) --- for which we can compute the corresponding average mass accretion efficiency based on single star mass-core mass relations (color-coded in Fig.~\ref{fig_gammacol}) --- whether the secondary would swell up and render the mass transfer unstable. For $\gamma\,$Columbae, this leads to the exclusion of the double-hatched area. An analysis of the possible accreted masses of pristine matter and matter in CNO-equilibrium can further constrain the early- and late-phase mass accretion efficiency (Methods). The result is that, if $\gamma\,$Columbae evolved through Case\,B, it experienced very inefficient mass transfer from a donor star of at least $15\mso$. \rev{See Extended Data Fig.~\ref{fig_uncertain} to see how the uncertainties in surface abundances affect the results.}

With the same method, we have analyzed the three stars HD\,48279 (ON8.5), HD\,93840 (BN1b), and $\zeta\,$Ophiuchi (O9.5IVnn). For HD\,93840, two different empirical stellar parameter sets were suggested \cite{Wessmayer2024}, which we investigate both, but we find that binary evolution can only explain one of them. For this and the other three stars, the initial masses and the mass accretion efficiency are again significantly constrained (see Extended Data Table~\ref{tab_obs} and Extended Data Fig.~\ref{fig_m1m2}).

Our detailed binary models and the results from our simple model suggest that apparent single stars which show the signature of CN-cycling in their surface abundances are naturally formed as mass gainers in binary systems. This offers a simpler explanation of their chemical properties than some previously suggested alternatives. At the same time, the observed stars cover a larger parameter space than our models in the CNO- as well as in the HeN-diagram. This is not surprising, since several uncertain physics parameters have been fixed for the model grid calculation. This concerns in particular the mass accretion efficiency and the efficiency for thermohaline mixing. Our result advocates that these uncertainties may be strongly reduced through the analysis of a larger sample of stars with envelopes enriched by the CN-cycle.

\textbf{Distinguishing rotational mixing}

Our models show clearly distinguishable effects of single star and binary evolution: accretion in binaries increases N/C far more than rotational mixing alone in single stars for moderate N/O. This is because accretion-induced mixing mixes the envelope efficiently without bringing core material into the envelope. Very fast rotating single stars, where rotational mixing is strong, are expected to show high N/C \textit{and} high N/O, since the mixing prevents the formation of a stabilizing mean molecular weight gradient. In contrast, accretion occurs \textit{after} this gradient has developed, inhibiting core-envelope mixing, enhancing N/C while keeping N/O moderate.

\textbf{Mergers}

Stellar mergers involve complex physics, and few multi-D models for massive mergers have been constructed so far \cite{Glebbeek2013,Schneider2019,Wu2020}.
However, the envelopes of merger products are thought to consist of a mixture of the envelope material of both stars. In post main sequence mergers, the core of the lower mass component, which is still at hydrogen burning densities, is also expected to dissolve and mix in the envelope \cite{Justham2014,Menon2024}. As a consequence, the envelope composition may reflect the incorporation of large amounts of helium-rich material in CNO-equilibrium (cf. Extended Data \Fig{fig_schematic}). As the merger process occurs on the short dynamical timescale, the associated envelope mixing process may also happen quickly. In this case, the resulting surface abundances are expected to show strong helium enhancement in the HeN diagram, and to lie near the ``CNO-eq. + dilution''-line in the CNO diagram. The latter feature would make them distinguishable from mass gainers emerging from stable mass transfer. This behavior is indeed found in the post-main sequence merger models from \cite{Menon2024}, which lie very close to our ``CNO-eq. + dilution''-line (note that this is different from the ``ON-cycle'' boundary in their fig.\,2). We note, however, that slow envelope mixing after the merger can not fundamentally be ruled out, which could raise the N/C-surface value.
 
As a proof-of-concept, we apply our analysis method to SN\,1987A, which is suggested to represent the explosion of a post-main sequence merger product \cite{Podsiadlowski1992}. For this we rely on the chemical abundance analysis of the circumstellar rings of SN\,1987A \cite{Lundqvist1996}, which are expected to reflect the surface abundances of the supernova progenitor shortly before the explosion. We find that they indeed reflect diluted CNO-equlilibrium abundances, without a discernible contribution from the CN-cycle (Fig.\,\ref{fig_CNO2}). The high CNO-equilibrium dilution factor of $\fcno\approx0.50$ implies that $6\dots 8\mso$ of pristine matter are required to have been lost to explain such a high CNO-equilibrium enrichment, and that the accreted companion was a $\sim$2.5$\mso$ star (see Supplementary Information Section A for details). While allowing mixing of matter from the outer part of the helium core during the merger would allow a wider range of solutions \cite{Menon2017}, this is not required by the observed CNO and helium abundances. 

\textbf{Outlook}

Modern large scale surveys of massive stars as in\cite{SimonDiaz2014,Shenar2024,SimonDiaz2026}, and Gaia \cite{MuellerHorn2025} are opening new doors to assess the massive star content in our local environment. This allows for the first time to empirically determine the line in the Hertzsprung-Russell diagram where single stars end core hydrogen burning \cite{deBurgos2025,Lennon2025}. At the same time, we are faced with the paradigm to account for core hydrogen as well as core helium burning binary products on either side of this line \cite{Wang2020,Menon2024,Wang2024,Jin2025}. 

In this situation, the analysis of the surface chemical composition of stars may be the key to unlock the different evolutionary histories of stars in the Hertzsprung-Russell diagram, and to determine the relative contributions of the different evolutionary paths to a given location. Applying this to a larger population will enable a comprehensive understanding of mass transfer and merger efficiency across the initial binary parameter space. 

Evolution via stable mass transfer and the common envelope channel have been suggested to lead to merging compact objects and gravitational wave emission \cite{Kruckow2016,Heuvel2017}. Both paths are governed by how mass and angular momentum is redistributed and lost. Improving our understanding of these processes will ultimately sharpen the predictions for these pathways and allow for conclusive comparisons with the gravitational wave sources detected by LIGO.

\section*{Methods}

\vskip 0.3cm\noindent{\bf Deriving the equations for our diagnostic lines}\\
When considering the surface enrichment of stars, often several elements or isotopes are considered simultaneously in a diagnostic diagram. It is useful to constrain the parameter space in such diagrams through lines that define the allowed subspace, which we derive in the following. 
To do so, we assume that a pristine stellar envelope of mass $\Mi$ is polluted with enriched matter of mass $\Me$, resulting in a total envelope mass $\Menv=\Mi+\Me$, with $f=\Me/(\Mi+\Me)=\Me/\Menv$ defining the dilution factor. We consider a species with initial mass fraction $X_{\rm i}$ 
which is enriched or depleted through mixing with matter of mass $\Me$ and mass fraction $X_{\rm e}$. \edi{We use mass fractions in this section, but note that mass fractions can be replaced by number fractions for the ratios between two elements/isotopes.} Limiting lines are particularly useful if the nucleosynthetic enrichment (or depletion) of the considered species is monotonous and ends with a finite ``equilibrium'' mass fraction. 
After complete mixing, the surface mass fraction $X$ is 
\begin{equation}
X= (1-f)X_\mathrm{i} + f\,X_\mathrm{e} .
\label{eqm1}
\end{equation} 
To consider up to four different species simultaneously, we define their initial and enriched mass fractions as $a_{\rm i}, b_{\rm i}, c_{\rm i}, d_{\rm i}$ and  $a_{\rm e}, b_{\rm e}, c_{\rm e}, d_{\rm e}$, respectively, so that their surface mass fractions after mixing $a, b, c, d$ can be calculated according to Eq.\,\ref{eqm1}.

When in a diagnostic diagram the abundance ratio $y:=a/b$ is plotted against the ratio $x:=c/d$, lines according to the above specified mixing conditions can be defined with the help of the parameter $f$ as
\begin{equation}
x=\frac{(1-f)c_{\rm i}+f c_{\rm e}}{(1-f)d_{\rm i}+f d_{\rm e}}, \\
y=\frac{(1-f)a_{\rm i}+f a_{\rm e}}{(1-f)b_{\rm i}+f b_{\rm e}}
\label{eqm2} ,
\end{equation}
or directly as
\begin{equation}
y(x)= \left(\frac{a}{b}\right)_{\rm i} \left(\frac{a}{b}\right)_{\rm e}
   \frac{\left( \left(\frac{d}{a}\right)_{\rm e} - \left(\frac{d}{a}\right)_{\rm i} \right) x
     +  \left(\frac{c}{a}\right)_{\rm i} -  \left(\frac{c}{a}\right)_{\rm e} }
       {\left( \left(\frac{d}{b}\right)_{\rm e} - \left(\frac{d}{b}\right)_{\rm i} \right) x
     +  \left(\frac{c}{b}\right)_{\rm i} -  \left(\frac{c}{b}\right)_{\rm e} }
       \label{eqm3},
\end{equation}
where $\left(\frac{a}{b}\right)_\mathrm{i} = \frac{a_\mathrm{i}}{b_\mathrm{i}}$, and the same notation applies to the other ratios.
We apply this formalism to two diagnostic surface abundance planes---the HeN diagram (\Fig{fig_HeN}) and the CNO diagram (\Fig{fig_CNO2})---both of which trace products of the CNO-cycle. This formalism can also be applied to other elements and isotopes.

In the first case, we investigate the nitrogen versus helium abundances in massive main sequence stars. Here, the idea is that a core hydrogen burning star in a binary system accretes part of its companion's envelope. Its envelope after accretion is then composed of its original pristine envelope ($M_{\rm 2, env}$), a fraction of the donor's pristine envelope ($M^{\rm acc}_{\rm 1, i}$), and of the material from the H/He-gradient region left by the receding convective core in the donor star ($M^{\rm acc}_{\rm 1, e}$), for which the CNO nuclei are in CNO-equilibrium. We assume that these three regions are mixed in the envelope of the mass gainer (cf. \Fig{fig_pie} and \,Extended Data Fig.~\ref{fig_schematic}). Then $\Mi=M_{\rm 2, env} + M^{\rm acc}_{\rm 1, i}$ and $\Me=M^{\rm acc}_{\rm 1, e}$, and the dilution factor is $\fcno = M^{\rm acc}_{\rm 1, e}/(M_{\rm 2, env} + M^{\rm acc}_{\rm 1, i}+M^{\rm acc}_{\rm 1, e})$. 

With this we can define lines in the HeN diagram according to Eq.\,\ref{eqm1} as
\begin{equation}
Y = (1-\fcno)Y_\mathrm{i} + \fcno\,\Ycno,\\ \hspace{1em}
N = (1-\fcno)N_\mathrm{i} + \fcno\,N_\mathrm{e}.
\label{eq_YNfcno}
\end{equation}
Here, $Y_{\rm i}$ and $Y_\mathrm{CNO}$ correspond to the initial helium mass fraction and the average helium mass fraction in the accreted CNO-equilibrium matter, and $N_\mathrm{i}$ and $N_\mathrm{e}$ are the initial and CNO-equilibrium nitrogen mass fractions. $Y$ and $N$ are the mass fractions of helium and nitrogen after accretion and mixing in the envelope. $\frac{N}{N_\mathrm{i}}\left(\fcno\right)$ correspond to the red lines in \Fig{fig_HeN} for various values of $\fcno$. Analogously to Eq.\,\ref{eqm3}, with $b_{\rm e}=b_{\rm i}$, $d_{\rm e}=d_{\rm i}$, $b_{\rm i}/d_{\rm i}=1$, $a_\mathrm{i}=N_\mathrm{i}$, $a_\mathrm{e}=N_\mathrm{e}$, $c_\mathrm{i}=Y_\mathrm{i}$, and $c_\mathrm{e}=Y_\mathrm{CNO}$, we obtain
\begin{equation}
N(Y)=\frac{(N_{\rm i}-N_\mathrm{e})Y + Y_{\rm i}N_\mathrm{e}-N_{\rm i}\Ycno}{Y_{\rm i}-\Ycno},
\label{eq_NY}
\end{equation}
which correspond to the blue lines in \Fig{fig_HeN} for various values of $\Ycno$.
The two lower dashed lines in fig.\,C.1 of \cite{Carlos2025} are constructed under the same assumptions for $\Ycno = 0.98$ and $\Ycno = 0.63$, with various values of $\fcno$ indicated by the red dots.

Considering the N/C and N/O mass fraction ratios simultaneously in one diagnostic diagram allows to differentiate the contributions from the CN-cycle and from the full CNO-cycle, which is not possible in the HeN diagram. This ansatz makes sense, because in the convective core, the CN-cycle reaches CN-equilibrium on the short thermal timescale of the star, and even CNO-equilibrium is reached after only 10\% of the protons are converted to helium (see fig.\,2 of \cite{Arnould1999}). On the other hand, layers of massive stars above their convective core may still be hot enough to reach CN-equilibrium far too cold to reach CNO-equilibrium. 

To compute the mass fraction ratios $N/C$ and $N/O$ resulting from mixing matter in CNO-equilibrium with pristine matter, with $x=N/O$ and $y=N/C$, we use Eqs.\,\ref{eqm2} and\,\ref{eqm3} and obtain
\begin{equation}
x = \frac{(1-\fcno)N_{\rm i} + \fcno N_{\rm e}}{(1-\fcno)O_{\rm i} + \fcno O_{\rm e}}, \\ \hspace{1em}
y = \frac{(1-\fcno)N_{\rm i} + \fcno N_{\rm e}}{(1-\fcno)C_{\rm i} + \fcno C_{\rm e}}, 
\label{eqm6}
\end{equation}
and
\begin{equation}
y(x)= \frac{\left( \NO_{\rm e} - \NO_{\rm i} \right)x}
{\left( \CO_{\rm e} - \CO_{\rm i} \right)x
+ \NO_{\rm e} \CO_{\rm i} - 
 \NO_{\rm i} \CO_{\rm e},
} 
 \label{eq_cno}
\end{equation}
where the parameters with subscript ``i'' refer to initial values, and those with ``e'' to CNO-equilibrium values. This defines the ``CNO-eq. + dilution''-line in \Fig{fig_CNO2}.
The slope of the line connecting (0,0) and $(x, y)$ in Eq.~\ref{eqm6} is 
\begin{equation}
\left(\frac{y}{x}\right) (f) = \frac{(1-\fcno)O_{\rm i}+\fcno O_{\rm e}}{(1-\fcno)C_{\rm i}+\fcno C_{\rm e}}.
 \label{eqm8}
\end{equation}
When $\fcno\ll1$,
the right-hand side of the above equation simplifies to $\OC_{\rm i}$; that is, the $(x,y)$ points lie on the straight line defined by
\begin{equation}
y(x)=\OC_{\rm i}x .  
\label{eqm8a}
\end{equation}
We find that this serves as a good approximation for $y(x)$ in Eq.~\ref{eq_cno} when $x$ is not close to the equilibrium value. In logarithmic scales as in \Fig{fig_CNO2}, the slope of the line in Eq.~\ref{eqm8a} is unity, with the initial CNO abundances determining the starting point.

In the mixture of CN-equilibrium matter with pristine matter, the oxygen abundance does not change ($C_\mathrm{i} + N_\mathrm{i} = C_\mathrm{e} + N_\mathrm{e}$), and therefore we obtain 
\begin{equation}
y(x) = \frac{x}{\CO_{\rm i} + \NO_{\rm i}  - x},
\label{eq_cn}
\end{equation}
which defines the ``CN-eq. + dilution''-line in \Fig{fig_CNO2}. 

Equations for limiting lines in the CNO-diagram have been previously derived in \cite{Maeder2014}, which are widely used in the literature. While their ansatz refers to time-evolution of the nucleosynthsis during CN/CNO-cycling, their equation for the limiting line for CN-cycling is identical to the one derived above. However, their line for CNO-cycling is not, with an increasing deviation for larger N/O-ratios (gray dotted line in Fig.~\ref{fig_CNO2}). This is expected, as in their derivation for this line it is assumed that the carbon abundance is constant, which is not realized when the full CNO-cycle is operating. While \cite{Maeder2014} note that their analytic solutions only apply for small deviations from the initial abundances, they are often used more widely in the literature. \edi{(Note that there is an error in the caption of their fig.\,1. The lower dotted line should correspond to their eq. (14), which is derived assuming C is constant, and the upper dotted line corresponds to their eq. (17),  which is derived assuming O is constant.)} On the other hand, the limiting lines derived here are identical to those shown in \cite{Langer2012}.

When the mixture of CNO-equilibrium matter with pristine matter, characterized by a dilution factor of $\fcno$, undergoes CN-cycling, we obtain
\begin{equation}
y(x) = \frac{x}{\frac{(1-\fcno)\Ci+\fcno\Ce}{(1-\fcno)\Oi+\fcno\Oe}+\frac{(1-\fcno)\Ni+\fcno\Ne}{(1-\fcno)\Oi+\fcno\Oe}-x},
\label{eq_fcno_cn}
\end{equation}
which corresponds to the dashed lines in \Fig{fig_CNO2}.
The CN-cycle decreases the carbon abundance until it reaches CN-equilibrium value, $\mathrm{C_{CN-eq.}}$. The nitrogen abundance increases by the amount decreased in carbon, while the oxygen abundance remaining constant. For any $\fcno$, the material that has reached CN-equilibrium can be described by
\begin{align}
\mathrm{C}=\mathrm{C_{CN-eq.}},\\
\mathrm{N = (1-\fcno)\,(C_i+N_i) + \fcno \, (C_e+N_e) - \mathrm{C_{CN-eq.}}},\\
\mathrm{O=(1-\fcno)\, O_i+\fcno \, O_e}. 
\end{align}
Using the condition $\mathrm{C_i+N_i+O_i=C_e+N_e+O_e}$, and eliminating the parameter $\fcno$ to express the relations in terms of $x$ and $y$, this reduces to 
\begin{equation}
y(x) = \frac{(\Ci+\Ni+\Oi-\mathrm{C_{CN-eq.}})\,x}{\mathrm{C_{CN-eq.}}\,(x+1)},
\label{eq_max}
\end{equation}
which is shown as the ``max'' line in \Fig{fig_CNO2}.

\vskip 0.3cm\noindent{\bf The analytic framework and its application}

\textbf{The analytic model.} Here we provide a detailed description of the simple analytic model that was introduced in the main text. We consider the structure of the primary at core hydrogen exhaustion and assume that the entire H/He gradient layer and the pristine envelope are transferred to the secondary, the masses of which  ($M_\mathrm{1,CNO}$ and $M_\mathrm{1,non-CNO}$, respectively) are functions of the initial primary mass. Thus, we are implicitly considering Case\,B mass transfer, which is
the most common mass transfer case. In the following, we explain in more detail how we infer the initial binary configuration and mass accretion efficiency.

\textit{Evolutionary mass.}
We estimate the evolutionary masses ($M_\mathrm{evol}$) of the observed stars based on their effective temperature ($\log \Teff$) and luminosity ($\log L$). We use single star evolutionary tracks from \cite{Jin2024} as a reference. We assume that the stars are undergoing core hydrogen burning (see Supplementary Fig.~\ref{fig_HRD}). We account for the overluminosity ($\Delta \log L$) due to helium enrichment using the relation $\Delta \log L = (\mu/\mu_0)^\gamma$, where $\mu$ is the mean molecular weight and $\mu_0=0.61$ is the initial value. We adopt a value of $\gamma=4$ for HD\,48279, $\gamma$ Columbae, $\zeta$ Ophiuchi, and the Mock star, and $\gamma=3$ for HD\,93840, according to fig.\,17 of \cite{Kohler2015}. We also account for the rotation of the observed stars and use the tracks of stellar models which rotate at the rate of $\vsini/0.78$, where $\vsini$ is the projected rotational velocity of the observed star and 0.78 is the average $\sin i$ considering randomly oriented rotational axes. 

\textit{Envelope mass.}
We assume full rejuvenation and adopt the envelope mass ($M_\mathrm{env}$) of the corresponding single star model with the same evolutionary mass ($M_\mathrm{evol}$).

\textit{Spectroscopic mass.}
The spectroscopic mass ($M_\mathrm{spec}$) is calculated using the relation $\log \frac{M_\mathrm{spec}}{\mso}=\log \frac{L}{\lso}-\log \frac{\mathcal{L}}{\mathcal{L_\odot}}$, where the spectroscopic luminosity $\mathcal{L}$ is obtained from $\log \frac{\mathcal{L}}{\mathcal{L_\odot}}=4\log \Teff \, \mathrm{[K]}-\logg \, \mathrm{[cm/s^2]}-10.61$, where $g$ is the surface gravity \cite{Langer2014}. 

\textit{Amount of accreted matter in CNO-equilibrium.}
The mass of CNO-equilibrium matter accreted by the star ($\Macccno$) is estimated as $\Macccno=\fcno \, \Menv$, while the rest of the envelope mass represents matter in non-CNO-equilibrium, as $M_\mathrm{2,env,non-CNO}=\Menv - \Macccno$.

\textit{Constraints on the initial primary mass ($\Mpi$).}
the mass of available CNO-equilibrium matter that is transferred is the mass of the H/He gradient layer ($M_\mathrm{1,CNO}$) in the primary at core hydrogen exhaustion. This mass increases with initial mass, unless the stellar wind is very strong. This gives our first constraint on the initial primary mass ($\Mpi$), such that $M_\mathrm{1,CNO}(\Mpi)>\Macccno$.

For a given accreted amount of mass $\Macccno$ that originates from the H/He gradient layer, the deeper it is from, the larger $\Ycno$ it entails. The mass and slope of the H/He gradient become larger and less steep, respectively, with initial mass. Thus, for large $\Macccno$ and $\Ycno$, it is possible that primaries with low $\Mpi$ do not contain a compatible H/He gradient layer. We can derive a constraint on the mass of the H/He gradient layer as $M_\mathrm{1,CNO} > \frac{\Macccno\,(1-\Yi)}{2\min(\Ycno-\Yi, \, 1-\Ycno)}$. Here we assumed that the helium mass fraction within the H/He gradient layer decreases linearly with mass from 1 to $\Yi$. Then the helium mass fraction of the deepest layer from which the accreted matter originates from is $\Ycno+ \frac{(1-Y_\mathrm{i})\,\Macccno}{2\,M_\mathrm{1,CNO}}$, while that of the shallowest layer is $\Ycno - \frac{(1-Y_\mathrm{i})\,\Macccno}{2\,M_\mathrm{1,CNO}}$ \rev{(see Extended Data Fig.~\ref{fig_Yderivation})}. Since the helium mass fraction in the H/He gradient region must lie within the range $(Y_\mathrm{i}, 1)$, this leads to the above inequality.

The above two criteria ensure that the primary's H/He gradient layer contains sufficient helium-rich, CNO-equilibrium material to reproduce the accreted mass and composition. This determines the minimum initial primary mass ($M_\mathrm{1,preMT,min}$).

\textit{Constraints on the initial secondary mass ($\Msi$).}
We consider the case where the star did not accrete any non-CNO-equilibrium matter and investigate the following two constraints. First, the mass of non-CNO-equilibrium matter in the envelope, $M_\mathrm{env,non-CNO}$, represents the maximum envelope mass before mass transfer. This envelope mass can be translated into the initial mass and represents the maximum available initial secondary mass. Second, the mass of the star subtracted by the accreted CNO-equilibrium matter, $M_\mathrm{evol} - \Macccno$, provides another constraint to the initial secondary mass. The smaller of the two values is adopted as the maximum initial secondary mass ($M_\mathrm{2,preMT,max}$).

\textit{Constraints from mass accretion efficiencies.}
We define the early-phase mass accretion efficiency, $\beta_\mathrm{early}$, as the ratio of the mass of non-CNO-equilibrium material accreted onto the secondary to the mass transferred by the primary, which is the entire envelope mass of the primary: $\beta_\mathrm{early}=M_\mathrm{acc,non-CNO}/M_\mathrm{1,non-CNO}$. Likewise, the late-phase mass accretion efficiency, $\beta_\mathrm{late}=\Macccno/M_\mathrm{1,CNO}$, refers to the accreted fraction of the transferred CNO-equilibrium material $M_\mathrm{1,CNO}$, which encompasses the entire H/He gradient region mass of the primary. Thus, a given $\Mpi$ determines the denominators of $\beta_\mathrm{early}$ and $\beta_\mathrm{late}$. A given $\Msi$ determines the total accreted mass, $M_\mathrm{acc}=M_\mathrm{evol}-\Msi$, and the non-CNO equilibrium accreted mass, $M_\mathrm{acc,non-CNO}=M_\mathrm{acc}-\Macccno=M_\mathrm{evol}-\Msi-\Macccno$, which enters $\beta_\mathrm{early}$ as the numerator. This is why $\beta_\mathrm{late}$ only depends on $\Mpi$, while $\beta_\mathrm{early}$ depends on both $\Mpi$ and $\Msi$ (see grey vertical lines and slanted white lines in \Fig{fig_gammacol}). The overall mass accretion efficiency is then $\beta=(M_\mathrm{acc,non-CNO}+\Macccno)/(M_\mathrm{1,non-CNO}+M_\mathrm{1,CNO})$. For a given pair of initial primary mass and initial secondary mass ($\Mpi, \Msi$), there corresponds a set of $\beta_\mathrm{early}$, $\beta_\mathrm{late}$, and $\beta$. If the initial primary mass is very high, the primary might have lost all pristine matter ($M_\mathrm{1,non-CNO}=0$), then the only available initial secondary mass is $\Msi=M_\mathrm{evol} - \Macccno$.

\textit{Constraints on initial mass ratios ($\qi$).}
For given values of $\beta_\mathrm{early}$ and $\beta_\mathrm{late}$, we evaluate the stability of mass transfer according to the results of \cite{Schuermann2024}, which do not depend on binary evolution models. We adopt the case corresponding to the assumption that the ejected material from the binary carries away the specific orbital angular momentum of the mass gainer. For a given mass accretion efficiency, we take the lowest initial mass ratio that allows for the stable mass transfer as the critical mass ratio, regardless of the initial orbital period. Thereby, we are implicitly considering the most generous case for stable mass transfer. 

For the early accretion phase with an accretion efficiency of $\beta_\mathrm{early}$, the mass ratio of $\Msi/\Mpi$ is considered. For the late accretion phase with an accretion efficiency of $\beta_\mathrm{late}$, the mass ratio of $(\Msi+M_\mathrm{acc,non-CNO})/(M_\mathrm{1,i}-M_\mathrm{1,non-CNO})$ is considered. This accounts for the fact that, by the time the primary has been stripped of its pristine envelope, its mass has been substantially decreased, increasing the mass ratio in favor of stable mass transfer.

At the high initial mass ratio end ($\qi>0.95$), the secondary evolves too rapidly to remain on the main sequence by the time the primary undergoes a supernova explosion. Such systems are therefore excluded, as the observed stars are mostly runaways.

\textbf{Application.} Extended Data Figure~\ref{fig_m1m2} shows the allowed ranges of initial primary and secondary masses for each observed star, along with the corresponding mass accretion efficiencies according to our simple model. Except for HD\,93840 (A), all stars have viable solutions. The lower part of Extended Data Table~\ref{tab_obs} summarizes the results. In the following, we discuss each star in detail.

HD\,48279 is claimed to be a binary interaction product, as it shows much stronger surface nitrogen enhancement than other stars in their sample with similar luminosity \cite{Martins2012,Markova2018}. As previously shown, its high N/C for their N/O and its strong helium enhancement for a moderate nitrogen enhancement also support this. A fairly high overall mass accretion efficiency of $\beta>0.25$ is required to reproduce the surface properties of HD\,48279. This star is a runaway star \cite{Gvaramadze2012,Burssens2020} without an indication of a companion (its radial velocity is constant) \cite{Mahy2009}. This suggests that the donor star likely exploded as a supernova, disrupting the binary and leaving the mass gainer as a runaway. Notably, HD\,48279 has been classified as O8.5 V zNstrvar? \cite{MaizApellaniz2013} (or alternatively, as ON7.5 V \cite{Mahy2009}), where the ``var'' label indicates spectral variability \cite{Sota2011}, which may be linked to the presence of a magnetic field. However, HD\,48279 is not formally classified as magnetic, as its measured field does not reach the $5-\sigma$ detection threshold \cite{Fossati2015}.

HD\,93840 is a runaway B-type supergiant with strong nitrogen enhancement. \cite{Wessmayer2024} invoked a binary evolution scenario, which involves multiple mass transfer events between the primary and secondary and the reverse supernova order (mass gainer explodes first). They suggested that it can be a progenitor of an imminent core collapse supernova to explain its unusually high N/C ratio for its N/O and overluminosity. The star's CNO abundance ratios align well with our mass gainer predictions. However, our mass gainer models during the core hydrogen and helium burning phases do not show significant overluminosity (cf. \cite{Sen2022}), and this might be due to the fact that the star has indeed evolved well beyond the core helium burning phase. If so, its evolutionary and envelope masses may be overestimated, allowing for a lower initial secondary mass. However, the values of $\fcno$ and $\Ycno$ inferred from the surface abundance planes remain unaffected by the star's evolutionary phase and still indicate a strong enrichment of helium-rich, CNO-equilibrium material. 

$\zeta$ Ophiuchi (HD\,149757) is a fast-rotating, runaway star, which shows strong helium and nitrogen enhancement---characteristics that have been attributed to a past mass accretion event \cite{Renzo2021}. This star shows $\fcno\sim0\%$, which implies a late-phase mass accretion efficiency of zero. This may suggest that rapid rotation hinders efficient accretion during the late-phase mass transfer. In addition, this star is known to exhibit various types of spectral and photometric variability, which might be linked to its magnetic field \cite{Hubrig2011b}.

\vskip 0.3cm\noindent{\bf Caveats}\\
Of the core hydrogen burning mass gainer models with $\logLLsun<5.7$, 81\% have experienced Case\,B mass transfer, 16\% Case\,A, and 3\% Case\,C. About 22\% of the Case\,B products undergo a second mass transfer \cite{Jin2025}. Therefore, with $\sim$63\% of all stable mass transfer events leading to a single Case\,B mass transfer, our simple model focuses on the most frequent channel contributing to the population of core hydrogen burning mass gainers. Different mass transfer cases involve donors at different evolutionary stages, therefore different core-envelope structures. However, as long as the transferred material consists of either pristine matter or hydrogen-burning products, our framework for inferring the amount of accreted CNO-equilibrium matter remains applicable, while different available relative amounts of pristine and CNO-equilibrium matter may shift the solution parameter space. For example, Case\,A primaries are expected to transfer a larger amount of pristine envelope material than Case\,B counterparts. This can be taken into account when constraining the initial primary mass. With such appropriate modifications, our simple model can be extended to incorporate other mass transfer cases.

In our simple model, we assume that the mass gainer has the same core-envelope structure as a single star model with the same evolutionary mass, i.e., we assume complete rejuvenation. The validity of this assumption depends on how efficiently the core grows in response to mass accretion. If core growth is inefficient, the envelope mass will be larger than in the case of complete rejuvenation, directly affecting the inference of the amount of accreted CNO-equilibrium matter and, consequently, other parameters relevant for reconstructing the accretion history.

During the mass accretion or merger phase, internal mixing may bring core material into the envelope and introduce an additional contribution of CNO-equilibrium and helium-rich matter to the envelope beyond that provided by the accreted matter itself. Such mixing has been invoked for the progenitor of SN\,1987A \cite{Menon2017}. However, for the stars considered in our study, no additional contribution is required to explain the observed properties.

The mass accretion efficiency in our mass gainer models is rotationally limited. Higher mass accretion efficiencies could increase the parameter space covered in the CNO- and HeN-diagrams (cf., fig.\,C1 of \cite{Carlos2025}). More efficient mass accretion during the late phase of mass transfer would allow more accretion of material from the donor's deep interior with high helium enrichment, which could lead to a more efficient thermohaline mixing and to higher surface N/C ratios. Notably, the determination of $\fcno$ and $\Ycno$ using the analytic lines derived in this work is not affected by uncertainties in mass accretion and thermohaline mixing efficiencies.

The initial abundances used in this work (Extended Data Table~\ref{tab_cno}) are solar abundances of \cite{Asplund2021}, which are also adopted in our binary models. The initial abundances and therefore abundance ratios of an observed stars may differ from solar values. Since the goal of this work is to introduce a novel method as a proof of concept, we do not attempt to constrain the initial abundances of individual objects, especially given the large uncertainties in the measured abundances \rev{(cf. Extended Data \Fig{fig_uncertain})}. However, we briefly note that $\gamma$ Columbae and $\zeta$ Ophiuchi are in the solar neighborhood ($<300$ pc), and the initial composition of early B-type stars in this region are known to be close to solar \cite{Nieva2012}. Therefore, such differences in the initial abundances are unlikely to affect our conclusion that their peculiar surface abundance patterns identify them as mass gainers and change their inferred values of $\fcno$ and $\Ycno$ significantly. We call for more systematic and precise abundance measurements to fully exploit the potential of our method.

Different rotational mixing treatments may be adopted in single star models. Yet, they are expected to show distinguishable surface abundance patterns from mass gainers. In Extended Data Fig.~\ref{fig_CNO_GENEVA}, models from \cite{Ekstrom2012} exhibit significantly stronger helium enhancement for moderate nitrogen enhancement compared to \cite{Jin2024} because chemical mixing is more strongly inhibited in the latter by mean molecular weight gradients, which affect helium more than nitrogen, suggesting that the HeN diagram may also provide an interesting testbed for the physics of rotational mixing. We see that models from \cite{Ekstrom2012} overlap considerably with our mass gainer models, which could render a comparison with observed stars ambiguous. But the inclusion of CNO measurement can break this degeneracy. In Extended Data Fig.~\ref{fig_CNO_GENEVA}, single star models from \cite{Jin2024} and \cite{Ekstrom2012} show a similar trend: N/C increases almost linearly with N/O on logarithmic scale. Mass gainers still show higher N/C values than single stars, regardless of rotational mixing treatment.

\rev{Finally, we can see from Fig.\,\ref{fig_CNO2} that our mass donor models do not fall exactly onto the ``CNO-eq.+dilution''-line, but are elevated above this line by a small amount. This occurs because shell hydrogen burning introduces some short-term convective and semiconvective mixing in the region above it, incorporating a fraction of the small CN-processing layer at the bottom of the donor envelope
into the H/He-gradient zone, which is later exposed during the mass transfer. For clarity, this effect has been neglected in the main discussion of this paper. }

\section*{Data Availability}
The input files, including initial abundances, opacity tables, nuclear networks, and zero-age main sequence models, necessary to reproduce the MESA calculations, are available along with the complete binary model grid at: \url{https://doi.org/10.5281/zenodo.18222252} \cite{Jinzenodo} and \url{https://wwwmpa.mpa-garching.mpg.de/stellgrid/}.

\section*{Acknowledgments}
We thank Alexander Heger, Chen Wang, Selma E. de Mink, and Götz Gräfener for their help. The authors gratefully acknowledge the granted access to the Bonna cluster hosted by the University of Bonn.

\section*{Funding Statement}
HJ received financial support for this research from the International Max Planck Research School (IMPRS) for Astronomy and Astrophysics at the Universities of Bonn and Cologne.

\section*{Author Contributions Statement}
HJ performed the stellar evolution calculations and conducted the analysis. HJ and NL developed the theoretical interpretation and interpreted the results. All authors discussed the results and contributed to the manuscript.

\section*{Competing Interests Statement}
The authors declare no competing interests.

\section*{Extended Data Figures / Tables}

\setcounter{figure}{0}
\renewcommand{\figurename}{Extended Data Figure}

\begin{figure*}
	\centering
	\includegraphics[width=0.83\linewidth]{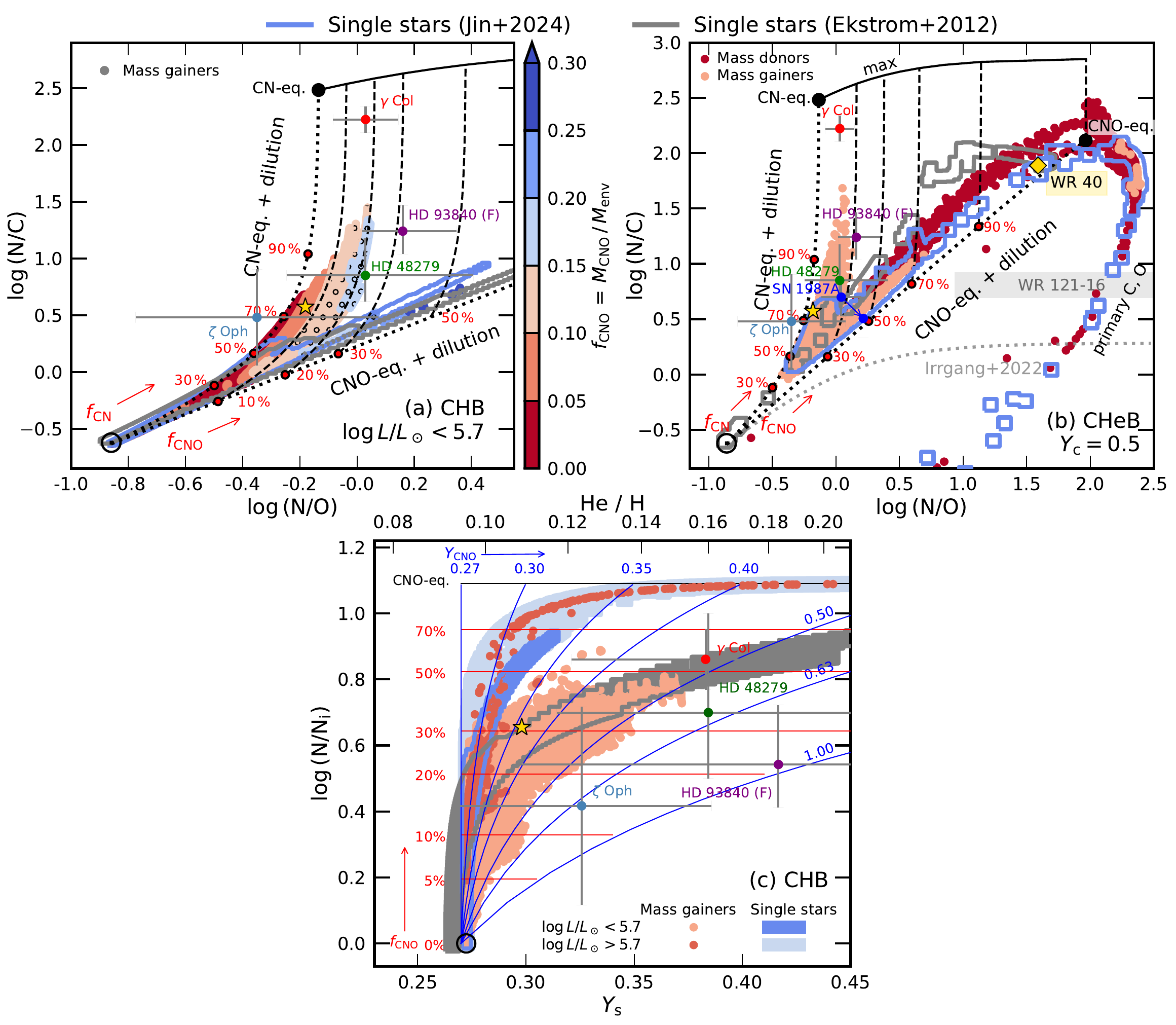}
	\caption{\textbf{Diagnostic surface abundance diagrams with different single star models.} Surface abundances of our mass gainer models (light and dark red), single star models (blue contours) from \cite{Jin2024}, and single star models from \cite{Ekstrom2012} (dark gray contours) with initial masses larger than $5 \mso$ and luminosities $\logLLsun < 5.7$ (with slightly different initial abundances compared to ours). The black lines in (a) and (b), blue and red lines in (c) represent analytic expressions for the composition of various mixtures (Eqs.\,\ref{eq_YNfcno}, \ref{eq_NY}, \ref{eq_cno}, \ref{eq_cn}, \ref{eq_fcno_cn}, \ref{eq_max}; see Methods). An empty black circle indicates the initial composition of our models. Several values of the dilution factors, $\fcn$ and $\fcno$, are marked by red circles and lines. The colored filled circles with error bars represent observed stars investigated in our study, where error bars denote 1$\sigma$ uncertainties, except for $\gamma$ Columbae (99\% confidence interval), and the yellow star symbol represents our ``Mock star''. (a): N/C and N/O abundance ratios at the surface during the core hydrogen burning phase. Colored filled circles represent mass gainer models every 100\,000 years after the accretion event, color-coded according to the $\fcno$ values determined from the models. Models with $\Ys > 0.33$ are marked with ``o''-hatching. The black filled circle corresponds to the pure equilibrium composition. (b): Same as (a), but here in the middle of the core helium burning phase for both, gainers and donors, without the color coding for the $\fcno$ values (cf. \Fig{fig_CNO2}). The yellow diamond marks the WN8 star WR\,40 \cite{Herald2001}, and the grey rectangle the WN/C star WR\,121-16 \cite{Zhang2020}. Note that the axis ranges are different from (a). (c): Helium mass fraction and nitrogen enhancement factor at the surface during the core hydrogen burning phase (cf. \Fig{fig_HeN}).}
	\label{fig_CNO_GENEVA}
\end{figure*}

\begin{figure}
\centering
\includegraphics[width=0.77\linewidth]{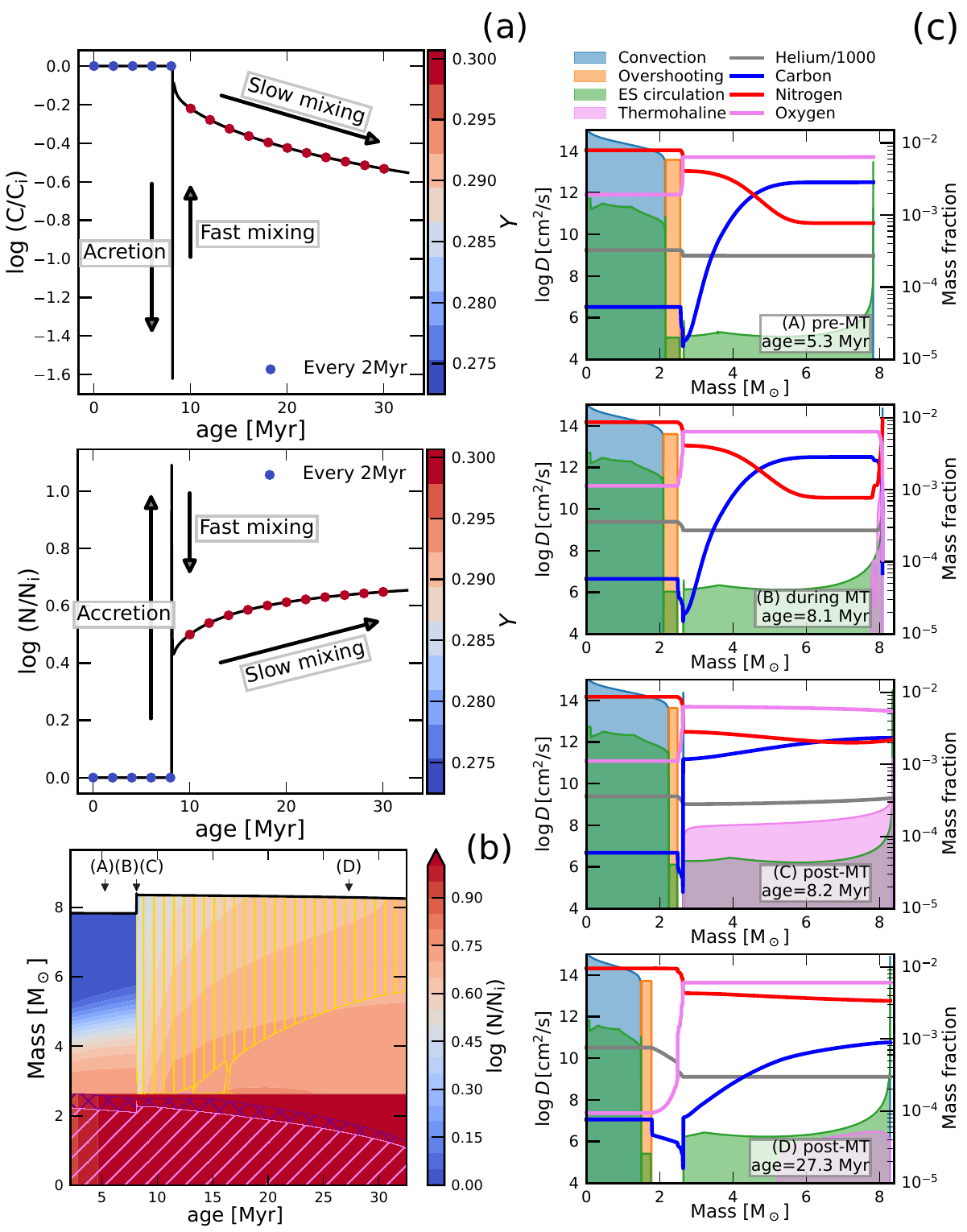}
\caption{{\bf Detailed chemical evolution of a mass gainer model.} The depicted model corresponds to our ``Mock star'' (see Extended Data Table\,\ref{tab_obs}). (a): Time evolution of the surface carbon depletion factor (top) and nitrogen enhancement factor (bottom). Colored dots are plotted every 2\,Myr, with the color corresponding to the surface helium mass fraction (see also fig.\,5 of \cite{Langer2010}). (b): Kippenhahn diagram showing the internal nitrogen enhancement factor (see color bar). Pink hatching indicates core convection, with purple hatching showing the core overshooting region. Yellow hatching indicates thermohaline mixing. (c): Internal profiles of diffusion coefficients for various mixing processes (left $y$-axis; see also fig.\,5 of \cite{Renzo2021}) and mass fractions of different chemicals (right $y$-axis) at four times, which are marked in the Kippenhahn diagram. The legend indicates the evolutionary phases. Phase (C) corresponds to ``Fast mixing'' and Phase (D) corresponds to ``Slow mixing'' indicated in \Fig{fig_pie} and panel (a). 
}
\label{fig_detail}
\end{figure}

\begin{figure}
	\centering
	\includegraphics[width=0.95\linewidth]{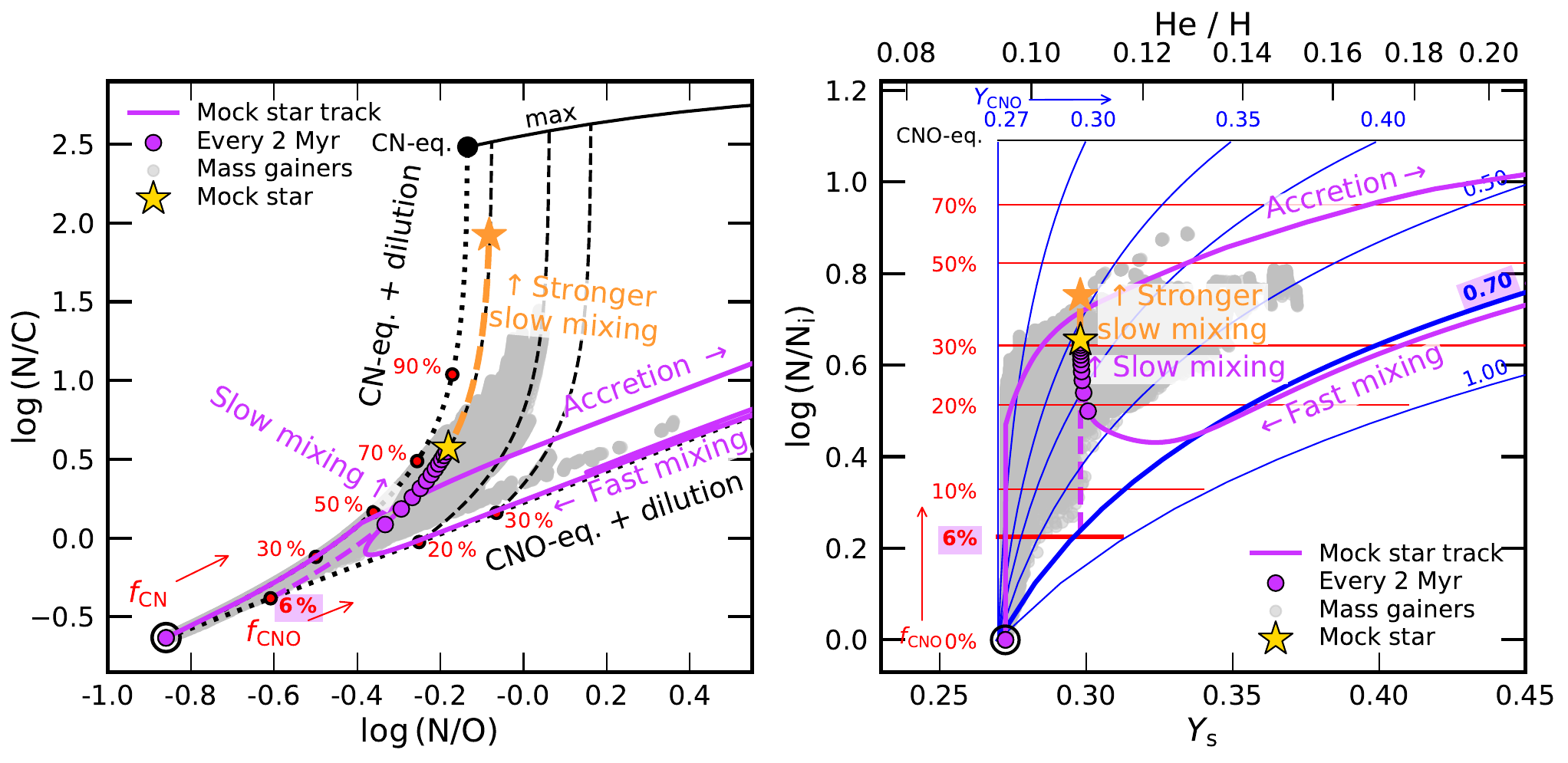}
	\caption{
    {\bf Evolutionary tracks of a mass gainer model in the diagnostic abundance planes.} The shown model is the one from which our ``Mock star'' (yellow star) is drawn. The purple solid lines are evolutionary tracks during the core hydrogen burning phase, with purple circles representing 2\,Myrs of evolution, plotted on top of other mass gainer models shown as grey circles. The left panel shows the CNO diagram, and the right panel the HeN diagram. The black lines (left) and blue and red lines (right) represent analytic expressions for the composition of various mixtures (Eqs.\,\ref{eq_YNfcno}, \ref{eq_NY}, \ref{eq_cno}, \ref{eq_cn}, \ref{eq_fcno_cn}, \ref{eq_max}; see Methods). The thick red and blue lines (right) correspond to the values for our ``Mock'' star. The purple dashed lines show how the $\fcno$ (CNO diagram; left) and $\Ycno$ (HeN diagram; right) are determined from the abundances of the Mock star. The orange dashed line and star show the effect of stronger slow mixing based on a newly computed model, wherein the efficiency of slow mixing was increased by a factor of 10 ($\alpha_\mathrm{th}=10$ for thermohaline mixing and $f_\mathrm{c}=0.3$ for rotational mixing). From the HeN diagram alone, only an upper limit of $\fcno<0.30$ can be obtained for the Mock star, while the CNO diagram constrains it to $\fcno=0.06$. Purple arrows indicate the effect of mass accretion, fast mixing, and slow mixing. An empty black circle indicates the initial composition of our models. Several values of the dilution factors, $\fcn$ and $\fcno$, are marked by red circles (left) and lines (right).
    }
	\label{fig_mock}
\end{figure}

\begin{figure}
\centering
\includegraphics[width=0.7\linewidth]{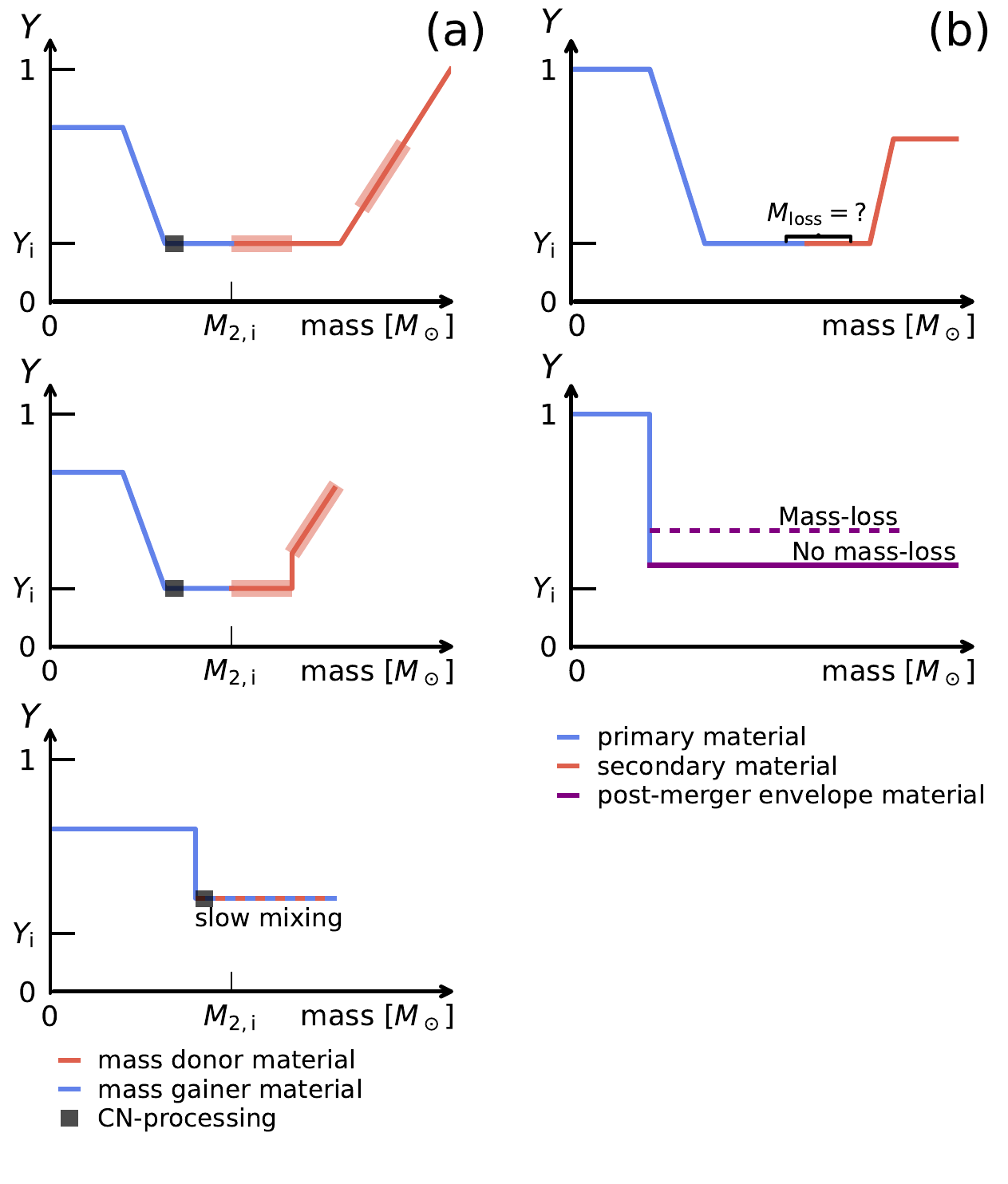}
\caption{{\bf Schematic internal helium profiles.} (a): Helium mass fraction as function of mass at three discrete stages in a mass transferring binary star (Case B mass transfer). Top panel: before mass accretion. Here, the mass gainers helium profile is depicted on the left side
(blue line, mass $< M_{\rm 2,i}$), and the helium profile of the transferred envelope of the donor star is added on the right side, with the layers that get transferred first closest to the surface layers of the accretor (mass = $M_{\rm 2,i}$). The highlighted parts of the donor profile indicate the parts of its envelope which, in this example, are assumed to be accreted. Middle panel: as the top panel, but with the non-accreted parts of the transferred donor envelope removed. Bottom panel: helium profile of the \rev{accretor} after its rejuvenation (core growth) and accretion-induced homogenization of the envelope, with the blue-red dashed part indicating where mass from both star is contributing. In all panels, the black squares indicate the layers undergoing CN-processing (b): Helium mass fraction $Y$ as function of mass before and after the merger of a primary star at the beginning of helium burning (blue line) with its core hydrogen burning companion star (red line). Complete envelope mixing is assumed. The cases without and with mass-loss during the merger event are depicted.
      }
	\label{fig_schematic}
\end{figure}

\begin{figure}
	\centering
	\includegraphics[width=0.8\linewidth]{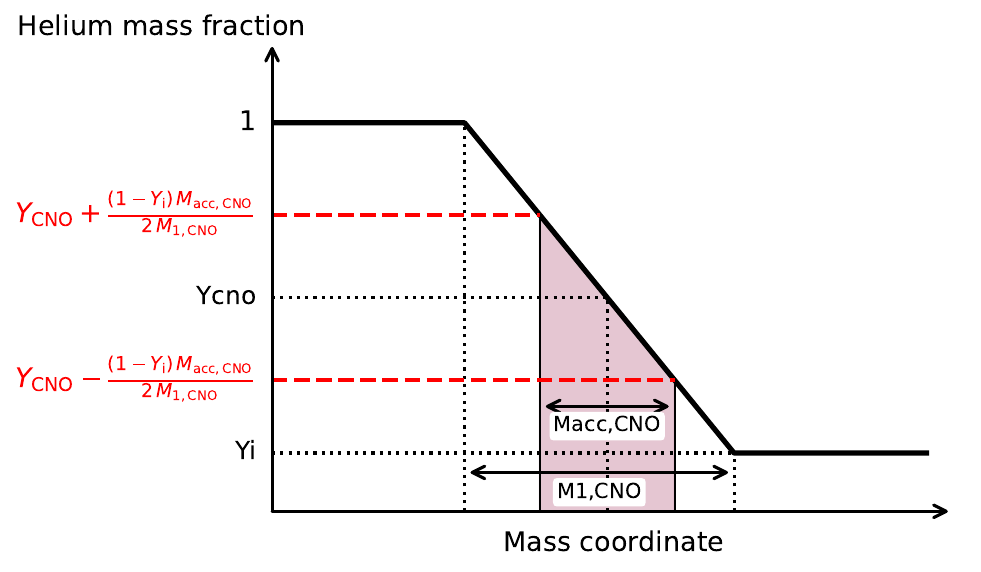}
	\caption{
    \rev{{\bf Schematic helium profile of a potential mass donor star.} It is shown for a time after the core hydrogen exhaustion, before the start of mass transfer. The variables have the same meanings as in Methods. The red shaded region indicates the part of the CNO-equilibrium matter which is accreted by the companion star. The helium mass fraction values marked in red are within the range of ($\Yi$, 1), which gives rise to the inequality $M_\mathrm{1,CNO} > \frac{\Macccno\,(1-\Yi)}{2\min(\Ycno-\Yi, \, 1-\Ycno)}$.}
    }
	\label{fig_Yderivation}
\end{figure}

\begin{figure}
	\centering
	\includegraphics[width=0.7\linewidth]{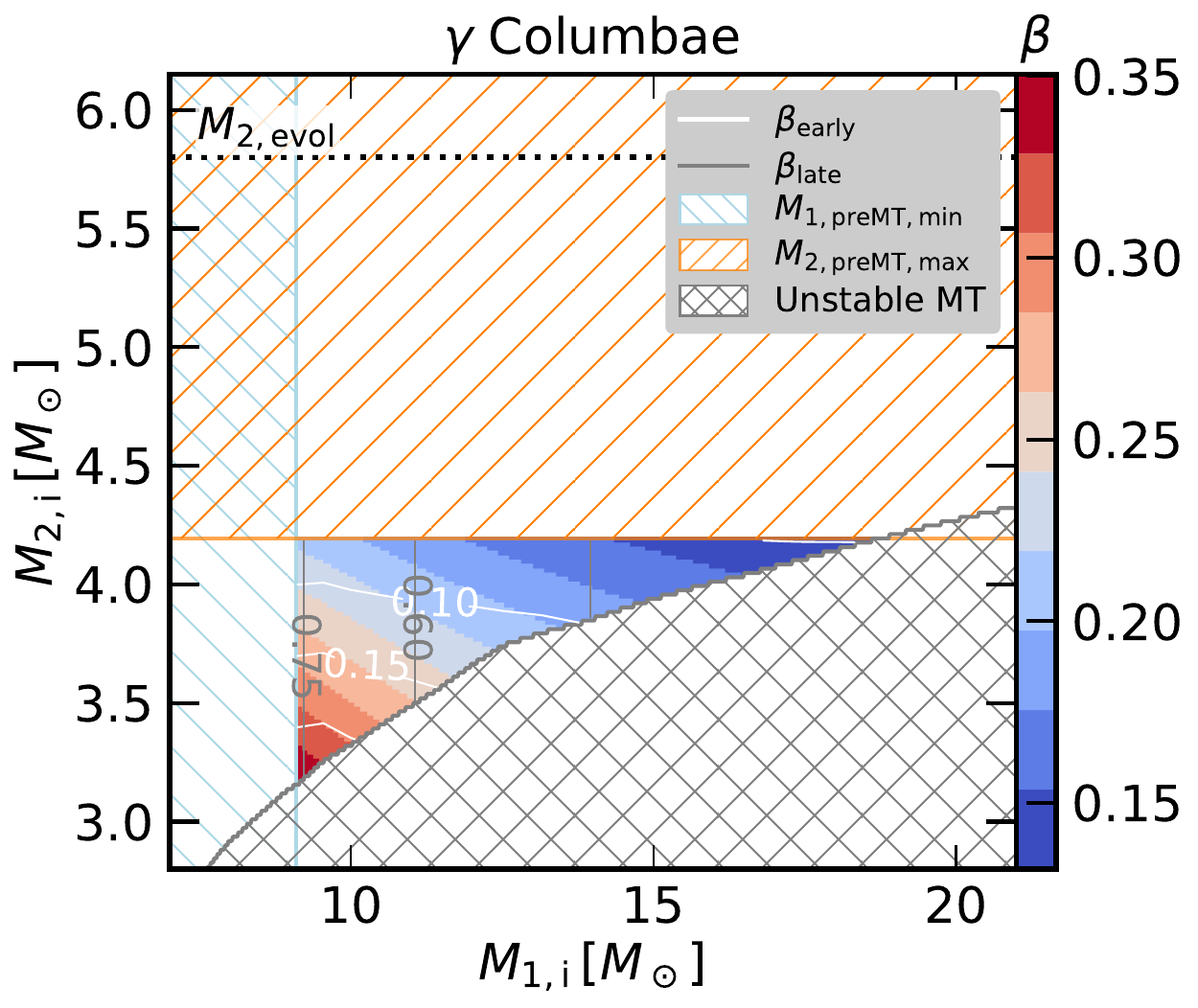}
	\caption{\rev{{\bf Initial masses diagram for $\gamma\,$Columbae for a higher N/O-value}. 
    The parameter space of the initial masses ($\Mpi$ and $\Msi$) of the components of the binary system that produced $\gamma\,$Col, when $\logNO=0.14$ is adopted instead of $\logNO=0.03$, which changes the allowed parameter space for the binary progenitor. Various constraints are indicated by different hatchings (see Methods). The colored region represents the possible initial configurations, where the color represents the average mass accretion efficiency ($\beta$) during the mass transfer. White and gray contour lines in this region denote allowed accretion efficiencies for the transfer of pristine ($\beta_{\rm early}$) and CNO-processed ($\beta_{\rm late}$) matter, respectively. The black horizontal dotted line indicates the current mass of $\gamma\,$Col.}
    }
	\label{fig_uncertain}
\end{figure}

\begin{figure}
	\centering
	\includegraphics[width=0.83\linewidth]{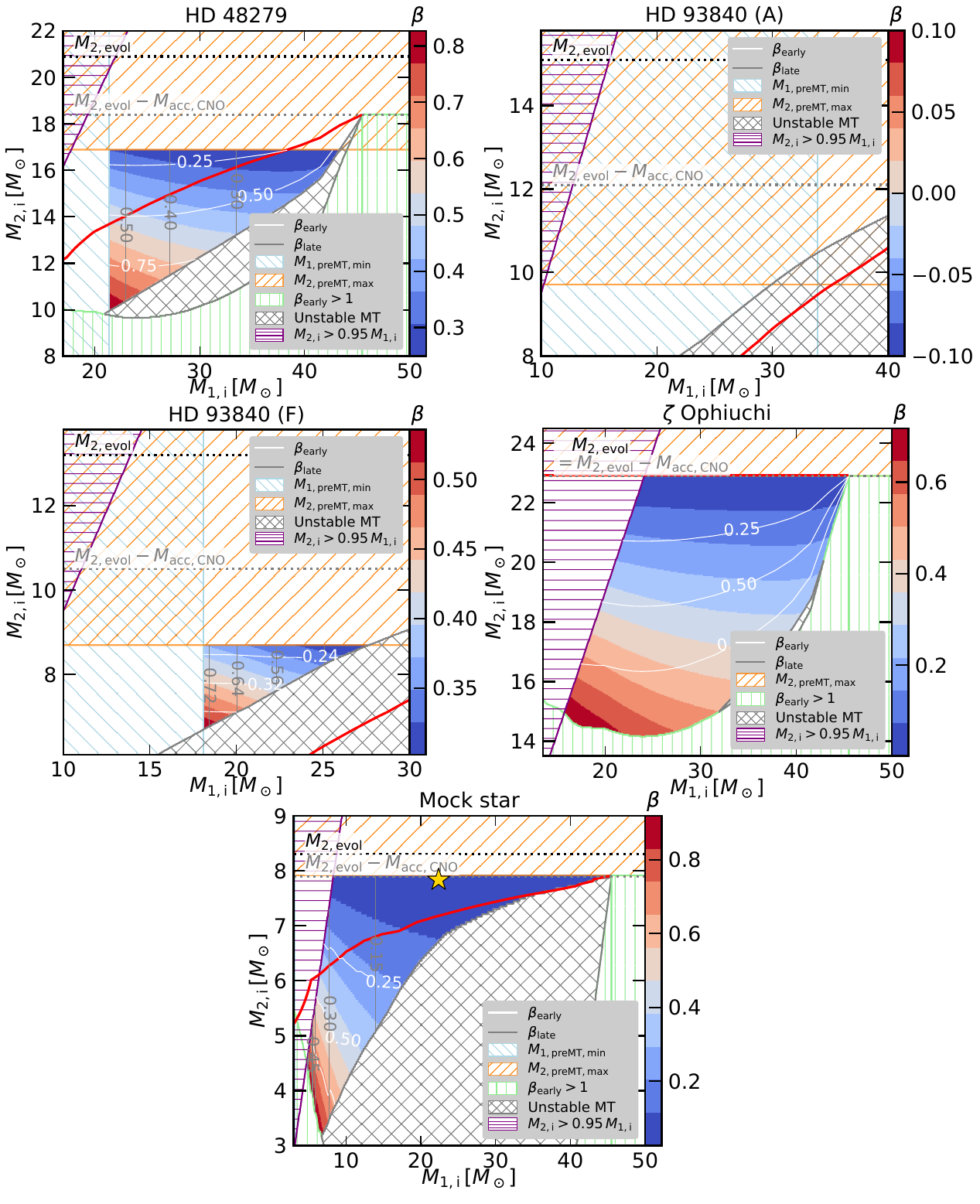}
	\caption{
    {\bf Possible initial configurations for the observed stars.}     
    Potential initial masses ($\Mpi$ and $\Msi$) of the components of the binary system that produced each of the five investigated OB star, where various constraints are indicated by different hatchings (see Methods). The colored region represents possible initial configurations, where the color represents the average mass accretion efficiency ($\beta$) during the mass transfer. White and gray contour lines in this region denote accretion efficiencies for the transfer of pristine ($\beta_{\rm early}$) and CNO-processed ($\beta_{\rm late}$) matter, respectively. The black horizontal dotted line indicates the current mass, and the grey horizontal dotted lines indicate the current mass subtracted by the accreted CNO-equilibrium mass. The thick red line marks where $\beta_\mathrm{early}$ and $\beta_\mathrm{late}$ are the same. The initial primary and secondary mass of the ``Mock star'' binary are marked with a star symbol.
    }
	\label{fig_m1m2}
\end{figure}

\renewcommand{\tablename}{Extended Data Table}
\begin{table}
    \caption{\textbf{Initial and equilibrium abundances used in this paper.} All values are mass fractions.}
    \centering
    \begin{tabular}{ccccc}
     \hline \hline
 & Initial & \begin{tabular}[c]{@{}c@{}}CN-eq.\\ ($17\,$MK)\end{tabular} & \begin{tabular}[c]{@{}c@{}}CNO-eq.\\ ($30\,$MK)\end{tabular} & \multicolumn{1}{c}{\begin{tabular}[c]{@{}c@{}}CNO-eq.\\ ($42\,$MK)\end{tabular}} \\ \hline
He & 2.72e-01 & - & - & - \\
C & 2.85e-03 & 1.18e-05 & 6.36e-05 & 1.01e-04 \\
N & 7.72e-04 & 4.09e-03 & 9.51e-03 & 9.51e-03 \\
O & 6.39e-03 & 6.38e-03 & 1.18e-04 & 6.71e-05 \\
N/C & 2.71e-01 & 3.48e+02 & 1.49e+02 & 9.43e+01 \\
N/O & 1.21e-01 & 6.42e-01 & 8.04e+01 & 1.42e+02 \\
C/O & 4.46e-01 & 1.84e-03 & 5.38e-01 & 1.50e+00 \\
     \hline
    \end{tabular}
    \label{tab_cno}
\end{table}

\renewcommand{\arraystretch}{0.85}
\begin{table}[]
\captionsetup{font=footnotesize}
\caption{\textbf{Empirical parameter of the selected stars.}
The definition of parameters can be found in the text. Rows $\fcno$...$\beta_\mathrm{late}$ present parameters derived from our framework. Only representative $\fcno$ values are used to derive the subsequent parameters. $^{(\ast)}$Parameters from both Atlas12+Detail+Surface (A) and FASTWIND (F) \cite{Wessmayer2024} $^{(^\P)}$Adopted initial abundances $^{(!)}$Peculiar velocities in case of runaways $^{(^\#)}$Conflicting claims of field detection $^{(\dagger)}$Values at $\beta=\beta_\mathrm{early}=\beta_\mathrm{late}$ $^{(\ddagger)}$Model data for the Mock star.}
\resizebox{\linewidth}{!}{%
\begin{tabular}{c|cccccc}
\hline \hline
 & HD 48279 & \begin{tabular}[c]{@{}c@{}}$\gamma$ Columbae\\ (HD 40494)\end{tabular} & \begin{tabular}[c]{@{}c@{}}HD 93840 (A)$^{\ast}$\\ HD 93840 (F)$^{\ast}$\end{tabular} & \begin{tabular}[c]{@{}c@{}}$\zeta$ Ophiuchi\\ (HD 149757)\end{tabular} & Mock star & Ref.$^{\P}$ \\ \hline
Spectral Type & \begin{tabular}[c]{@{}c@{}}O8.5 V zNstrvar? \cite{MaizApellaniz2013}\\ ON8.5 V \cite{Martins2015}\end{tabular} & B2.5IV \cite{Hiltner1969} & BN1 Ib \cite{Wessmayer2024} & O9.5IVnn \cite{Sota2014} & $-$ &  \\ \hline
$\logTeff$ & 4.54$_\mathrm{-0.01}^\mathrm{+0.01}$ \cite{Martins2015} & 4.19$_\mathrm{-0.01}^\mathrm{+0.01}$ \cite{Irrgang2022} & \begin{tabular}[c]{@{}c@{}}4.34$_\mathrm{-0.01}^\mathrm{+0.01}$ \cite{Wessmayer2024}\\ 4.36$_\mathrm{-0.02}^\mathrm{+0.02}$ \cite{Wessmayer2024}\end{tabular} & 4.53$_\mathrm{-0.03}^\mathrm{+0.03}$ \cite{Villamariz2005} & 4.27 &  \\ \hline
$\log L$ [$L_\odot$] & 4.95$_\mathrm{-0.11}^\mathrm{+0.11}$ \cite{Martins2012} & 3.53$_\mathrm{-0.12}^\mathrm{+0.12}$ \cite{Irrgang2022} & 4.71$_\mathrm{-0.08}^\mathrm{+0.08}$ \cite{Wessmayer2024} & 4.93$_\mathrm{-0.20}^\mathrm{+0.20}$ \cite{Villamariz2005} & 3.88 &  \\ \hline
$\log g$ [$\mathrm{cm^2/s}$] & 3.80$_\mathrm{-0.15}^\mathrm{+0.15}$ \cite{Martins2015} & 3.33$_\mathrm{-0.10}^\mathrm{+0.10}$ \cite{Irrgang2022} & \begin{tabular}[c]{@{}c@{}}3.00$_\mathrm{-0.05}^\mathrm{+0.05}$ \cite{Wessmayer2024}\\ 2.90$_\mathrm{-0.10}^\mathrm{+0.08}$ \cite{Wessmayer2024}\end{tabular} & 3.70$_\mathrm{-0.10}^\mathrm{+0.10}$ \cite{Villamariz2005} & 3.51 &  \\ \hline
$M_\mathrm{spec}$ [$M_\odot$] & 15.8$_\mathrm{-7.1}^\mathrm{+7.1}$ & 5.1$_\mathrm{-1.9}^\mathrm{+1.9}$ & \begin{tabular}[c]{@{}c@{}}9.1$_\mathrm{-2.1}^\mathrm{+2.1}$\\ 5.8$_\mathrm{-1.9}^\mathrm{+1.9}$\end{tabular} & 13.2$_\mathrm{-7.8}^\mathrm{+7.8}$ & 8.3 &  \\ \hline
$\Ys$ & 0.38$_\mathrm{-0.07}^\mathrm{+0.07}$ \cite{Martins2015} & 0.38$_\mathrm{-0.06}^\mathrm{+0.06}$ \cite{Irrgang2022} & \begin{tabular}[c]{@{}c@{}}0.31$_\mathrm{-0.04}^\mathrm{+0.04}$ \cite{Wessmayer2024}\\ 0.42$_\mathrm{-0.12}^\mathrm{+0.11}$ \cite{Wessmayer2024}\end{tabular} & 0.33$_\mathrm{-0.06}^\mathrm{+0.06}$ \cite{Villamariz2005} & 0.30 & 0.27 \\ \hline
$\log \mathrm{(N/O)}$ & $0.03_\mathrm{-0.28}^\mathrm{+0.38}$ \cite{Martins2015} & $0.03_\mathrm{-0.11}^\mathrm{+0.11}$ \cite{Irrgang2022} & \begin{tabular}[c]{@{}c@{}}$0.17_\mathrm{-0.09}^\mathrm{+0.09}$ \cite{Wessmayer2024}\\ $0.16_\mathrm{-0.14}^\mathrm{+0.19}$ \cite{Wessmayer2024}\end{tabular} & $-0.35_\mathrm{-0.42}^\mathrm{+0.42}$ \cite{Villamariz2005} & -0.18 & -0.86 \\ \hline
$\log \mathrm{(N/C)}$ & 0.85$_\mathrm{-0.23}^\mathrm{+0.33}$ \cite{Martins2015} & $2.22_\mathrm{-0.12}^\mathrm{+0.12}$ \cite{Irrgang2022} & \begin{tabular}[c]{@{}c@{}}$1.34_\mathrm{-0.12}^\mathrm{+0.12}$ \cite{Wessmayer2024}\\ $1.24_\mathrm{-0.20}^\mathrm{+0.23}$ \cite{Wessmayer2024}\end{tabular} & $0.48_\mathrm{-0.42}^\mathrm{+0.42}$ \cite{Villamariz2005} & 0.57 & -0.63 \\ \hline
$\log \mathrm{(N/N_i)}$ & $0.70_\mathrm{-0.20}^\mathrm{+0.30}$ \cite{Martins2015} & $0.86_\mathrm{-0.09}^\mathrm{+0.09}$ \cite{Irrgang2022} & \begin{tabular}[c]{@{}c@{}}$0.68_\mathrm{-0.08}^\mathrm{+0.08}$ \cite{Wessmayer2024}\\ $0.54_\mathrm{-0.13}^\mathrm{+0.18}$ \cite{Wessmayer2024}\end{tabular} & $0.42_\mathrm{-0.30}^\mathrm{+0.30}$ \cite{Villamariz2005} & 0.65 & 0.00 \\ \hline
$v \sin i$ [km/s] & 137$_\mathrm{-10}^\mathrm{+10}$ \cite{Martins2015} & $<$70 \cite{Irrgang2022} & 68 $_\mathrm{-3}^\mathrm{+3}$ \cite{Wessmayer2024} & $>$348 \cite{Gordon2018,Zehe2018} & 211 &  \\ \hline
Peculiar velocity$^{!}$ [km/s] & 28.6$_\mathrm{-5.4}^\mathrm{+5.4}$ \cite{Gvaramadze2012} & $-$ & $\sim 28$ \cite{Wessmayer2024} & 20-50 \cite{Zehe2018,Neuhaeuser2020} & 143 &  \\ \hline
Misc. & \begin{tabular}[c]{@{}c@{}}No field detection \cite{Fossati2015}\\ Sp. var. \cite{MaizApellaniz2013}\end{tabular} & \begin{tabular}[c]{@{}c@{}}Magnetic?$^{\#}$ \cite{Hubrig2009,Bagnulo2012}\\ Sp. var. \cite{Irrgang2022}\end{tabular} & Over. lum. \cite{Wessmayer2024} & Magnetic \cite{Hubrig2011b} & $-$ &  \\ \hline \hline
$\fcno$ & 22\%$_\mathrm{-22\%}^\mathrm{+35\%}$ & 17\%$_\mathrm{-12\%}^\mathrm{+12\%}$ & 32\%$_\mathrm{-10\%}^\mathrm{+10\%}$ & 0\%$_\mathrm{-0\%}^\mathrm{+35\%}$ & \begin{tabular}[c]{@{}c@{}}6\%\\ (6\%$^{\ddagger}$)\end{tabular} &  \\ \hline
$\Ycno$ & 0.8 & 0.9 & \begin{tabular}[c]{@{}c@{}}0.4\\ 0.7\end{tabular} & Any & \begin{tabular}[c]{@{}c@{}}0.7\\ (0.6$^{\ddagger}$)\end{tabular} &  \\ \hline
$\Delta \log L$ [$L_\odot$] & 0.18 & 0.18 & \begin{tabular}[c]{@{}c@{}}0.03\\ 0.18\end{tabular} & 0.09 & \begin{tabular}[c]{@{}c@{}}0.04\\ (0.03$^{\ddagger}$)\end{tabular} &  \\ \hline
$M_\mathrm{evol}$ [$M_\odot$] & 20.9$_\mathrm{-0.9}^\mathrm{+1.0}$ & 6.0$_\mathrm{-0.2}^\mathrm{+0.6}$ & \begin{tabular}[c]{@{}c@{}}15.1$_\mathrm{-1.3}^\mathrm{+0.7}$\\ 13.2$_\mathrm{-0.6}^\mathrm{+1.3}$\end{tabular} & 22.9$_\mathrm{-2.0}^\mathrm{+2.2}$ & \begin{tabular}[c]{@{}c@{}}8.3\\ (8.3$^{\ddagger}$)\end{tabular} &  \\ \hline
$\Menv$ [$M_\odot$] & 11.4$_\mathrm{-0.3}^\mathrm{+0.3}$ & 4.4$_\mathrm{-0.1}^\mathrm{+0.4}$ & \begin{tabular}[c]{@{}c@{}}9.3$_\mathrm{-0.6}^\mathrm{+0.3}$\\ 8.5$_\mathrm{-0.3}^\mathrm{+0.6}$\end{tabular} & 12.0$_\mathrm{-0.6}^\mathrm{+0.6}$ & \begin{tabular}[c]{@{}c@{}}5.9\\ (5.7$^{\ddagger}$)\end{tabular} &  \\ \hline
$M_\mathrm{acc,CNO}$ [$M_\odot$] & 2.5 & 0.8 & \begin{tabular}[c]{@{}c@{}}3.0\\ 2.7\end{tabular} & 0.0 & \begin{tabular}[c]{@{}c@{}}0.4\\ (0.3$^{\ddagger}$)\end{tabular} &  \\ \hline
$M_\mathrm{env,non-CNO}$ [$M_\odot$] & 8.9 & 3.7 & \begin{tabular}[c]{@{}c@{}}6.3\\ 5.8\end{tabular} & 12.0 & \begin{tabular}[c]{@{}c@{}}5.5\\ (5.4$^{\ddagger}$)\end{tabular} &  \\ \hline
$M_\mathrm{2,preMT,max}$ [$M_\odot$] & 16.9 & 5.0 & \begin{tabular}[c]{@{}c@{}}9.7\\ 8.7\end{tabular} & 22.9 & 7.9 &  \\ \hline
$M_\mathrm{1,preMT,min}$ [$M_\odot$] & 21.4 & 14.2 & \begin{tabular}[c]{@{}c@{}}33.9\\ 18.1\end{tabular} & $-$ & 3.1 &  \\ \hline
$M_\mathrm{1,i}$ [$M_\odot$] & \begin{tabular}[c]{@{}c@{}}21.4-43.3\\ (21.4-38.3$^{\dagger}$)\end{tabular} & 14.2-19.5 & \begin{tabular}[c]{@{}c@{}}No solution\\ 18.1-27.8\end{tabular} & \begin{tabular}[c]{@{}c@{}}15.8-45.5\\ (24.1-45.5$^{\dagger}$)\end{tabular} & \begin{tabular}[c]{@{}c@{}}4.8-45.2\\ (22.4$^{\ddagger}$)\end{tabular} &  \\ \hline
$M_\mathrm{2,i}$ [$M_\odot$] & \begin{tabular}[c]{@{}c@{}}10.0-16.9\\ (13.7-16.9$^{\dagger}$)\end{tabular} & 4.4-5.0 & \begin{tabular}[c]{@{}c@{}}No solution\\ 6.8-8.7\end{tabular} & \begin{tabular}[c]{@{}c@{}}14.2-22.9\\ (22.9-22.9$^{\dagger}$)\end{tabular} & \begin{tabular}[c]{@{}c@{}}3.2-7.9\\ (7.8$^{\ddagger}$)\end{tabular} &  \\ \hline
$q_\mathrm{i}$ & \begin{tabular}[c]{@{}c@{}}0.38-0.79\\ (0.44-0.64$^{\dagger}$)\end{tabular} & 0.26-0.35 & \begin{tabular}[c]{@{}c@{}}No solution\\ 0.31-0.48\end{tabular} & \begin{tabular}[c]{@{}c@{}}0.45-0.95\\ (0.53-0.95$^{\dagger}$)\end{tabular} & \begin{tabular}[c]{@{}c@{}}0.17-0.95\\ (0.35$^{\ddagger}$)\end{tabular} &  \\ \hline
$\beta$ & \begin{tabular}[c]{@{}c@{}}0.25-0.82\\ (0.25-0.55$^{\dagger}$)\end{tabular} & 0.08-0.16 & \begin{tabular}[c]{@{}c@{}}No solution\\ 0.30-0.53\end{tabular} & \begin{tabular}[c]{@{}c@{}}0.00-0.72\\ ($\sim0.00$$^{\dagger}$)\end{tabular} & \begin{tabular}[c]{@{}c@{}}0.03-0.92\\ (0.05$^{\ddagger}$)\end{tabular} &  \\ \hline
$\beta_\mathrm{early}$ & 0.17-0.97 & 0.02-0.12 & \begin{tabular}[c]{@{}c@{}}No solution\\ 0.21-0.44\end{tabular} & 0.00-1.00 & \begin{tabular}[c]{@{}c@{}}0.00-1.00\\ (0.03$^{\ddagger}$)\end{tabular} &  \\ \hline
$\beta_\mathrm{late}$ & 0.21-0.55 & 0.14-0.29 & \begin{tabular}[c]{@{}c@{}}No solution\\ 0.42-0.74\end{tabular} & 0.00-0.00 & \begin{tabular}[c]{@{}c@{}}0.03-0.52\\ (0.11$^{\ddagger}$)\end{tabular} &  \\ \hline
\end{tabular}
\label{tab_obs}
}
\end{table}

\newpage
\setcounter{page}{1}

\section*{Supplementary information}
\setcounter{figure}{0}
\renewcommand{\figurename}{Supplementary Figure}

\subsection*{A: A closer look at SN\,1987A}\label{app_87A}

\begin{figure}
	\centering
	\includegraphics[width=0.8\linewidth]{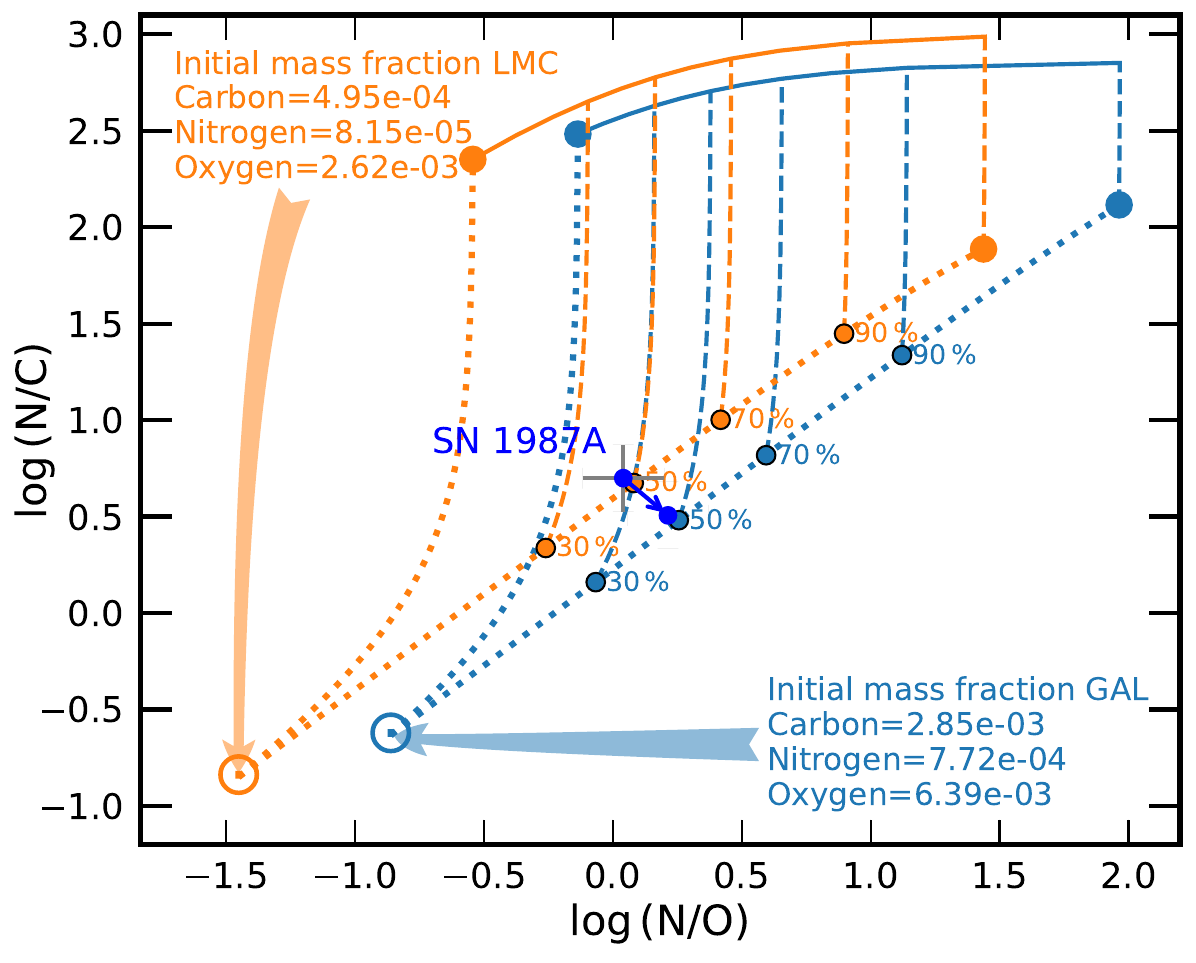}
    \caption{{\bf Diagnostic CNO-surface abundance diagram for two different initial compositions.}    
    \edi{The dotted lines represent analytic expressions for the composition of a mixture of pristine matter with matter containing CN- (Eq.\,\ref{eq_cn}), or CNO-equilibrium abundances (Eq.\,\ref{eq_cno}),} derived from two sets of initial CNO abundances: the Solar neighborhood \cite{Jin2025} (blue; this work) and the Large Magellanic Cloud \cite{Brott2011} (orange) (see Methods). \edi{Empty circles indicate the initial CNO abundances, and filled circles at the upper right ends of the dotted lines indicate the pure equilibrium compositions. Several values of the dilution factor $\fcno$ are marked by circles along the bottom dotted lines. The dashed lines equidistant to the ``CN-eq. + dilution''-line are obtained by assuming CN-processing, starting with a composition at the ``CNO-eq. + dilution''-line (Eq.\,\ref{eq_fcno_cn}). They end at the top full-drawn line, which represents complete CN-processing (Eq.\,\ref{eq_max}). The blue filled circle with error bars represent observed values for SN\,1987A \cite{Lundqvist1996}, where error bars denote 1$\sigma$ uncertainties.} The blue arrow shows the required correction to interpret the abundances of the SN\,1987A progenitor in a diagnostic diagram with Solar abundances.
    }
	\label{fig_LMC}
\end{figure}
    
\begin{figure}
	\centering
	\includegraphics[width=0.95\linewidth]{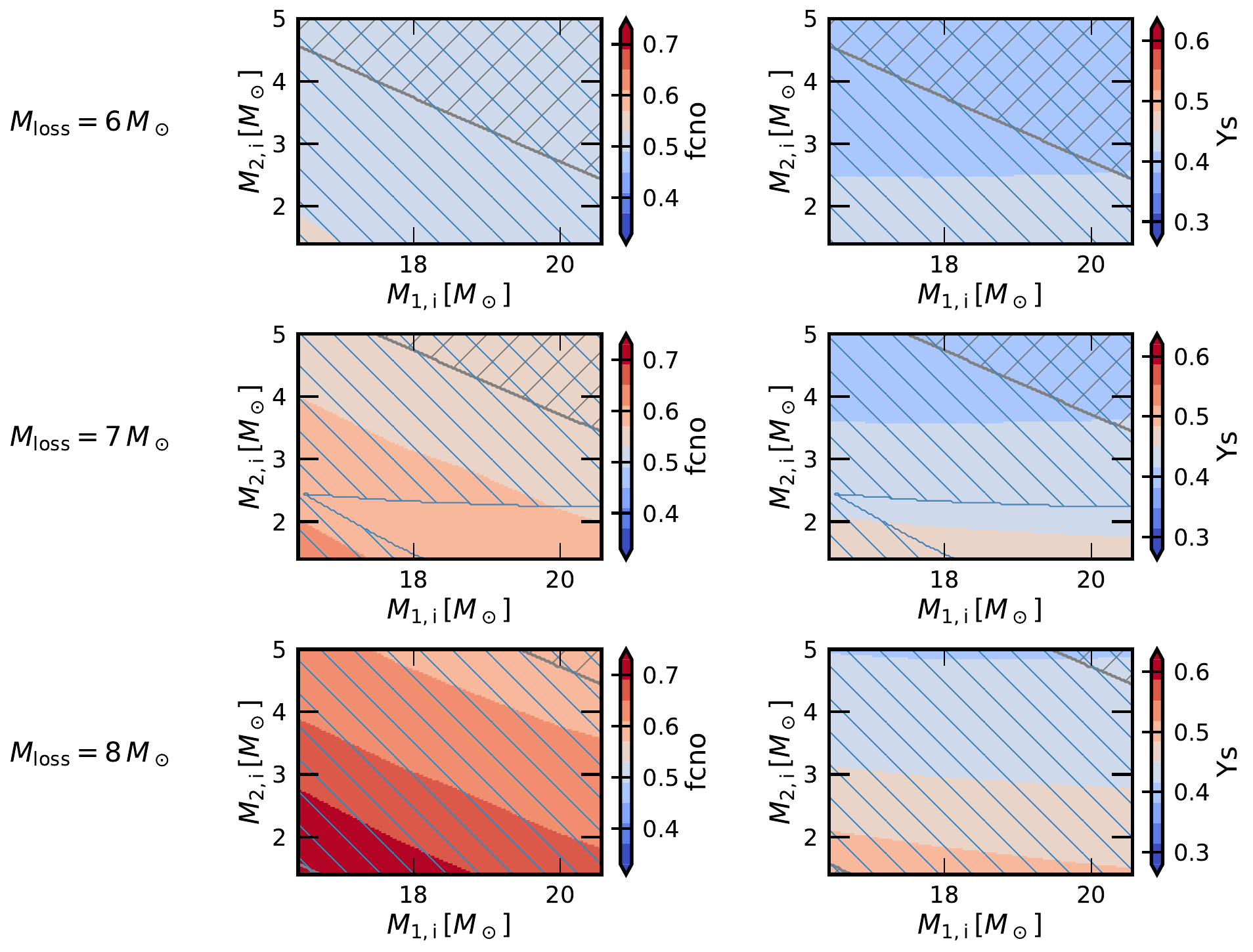}
	\caption{{\bf Possible initial configurations for the progenitor of SN\,1987A.} The plotted $\Mpi$ range correspond to helium core masses falling within $5-7\Msun$, and regions where hydrogen envelope mass falls outside $5-10\mso$ are marked with ``//'', constraints from supernova models \cite{Woosley1988}. Regions where the predicted $\fcno$ and $\Ys$ values deviate significantly from the estimated values ($\fcno$ by more than 10\% from 50\% and $\Ys$ more than 0.05 from 0.49, respectively) are marked with ``$\backslash\backslash$''.
    }
	\label{fig_87A}
\end{figure}

Spectral analysis of circumstellar rings around SN \,1987A has revealed abundance ratios of N/C=$5.0\pm2.0$ and N/O=$1.1\pm0.4$ (cf. Fig.~\ref{fig_CNO2}) and He/H=$0.25\pm0.05$ (helium mass fraction of $\sim0.44-0.54$) \cite{Lundqvist1996}. 
We assess these abundances in a similar manner as in mass gainers to constrain the pre-merger configuration assuming (a) the chemical composition of the envelope of the merger product is the result of a fast mixing of the envelope of the primary and the entirety of the secondary, while the helium core of the merger product is the helium core of the primary (cf. Extended Data Fig.~\ref{fig_schematic}), and (b) the merger product has a hydrogen-rich envelope mass of $5-10\mso$ and a helium core mass of $5-7\mso$, as inferred from supernova modeling \cite{Woosley1988}. The amount of mass lost during the merger ($M_\mathrm{loss}$) is set as a free parameter. 

As discussed in Fig.~\ref{fig_CNO2}, shifting the observed position of SN\,1987A in the CNO diagram to correct for the non-solar CNO-ratios in the Large Magellanic Cloud makes it fall straight onto the ``CNO-eq. + dilution''-line, with no CN-cycling component (Supplementary Fig.~\ref{fig_LMC}). This supports our assumption that the envelope of the merger product was mixed quickly to achieve a homogeneous chemical composition, and that no long-term mixing process operated afterwards. Using the present day chemical composition of the Large Magellanic Cloud as in \cite{Brott2011}, we find that the progenitor's envelope was highly enriched with CNO-equilibrium matter, with $\fcno=50\%$.

In Supplementary Fig.~\ref{fig_87A}, we show the allowed parameter space as in Extended Data Fig.~\ref{fig_m1m2}, varying the amount of pristine envelope material lost in the merger process. While we find no solutions for $M_\mathrm{loss}\leq 6\mso$ and for $M_\mathrm{loss}\geq 8\mso$, we find that for $M_\mathrm{loss}\simeq 7\mso$, primaries with initial masses above $\sim 17\mso$ merging with a star of less than $\sim 2.5\mso$ can fulfill all constraints,
i.e., reproduce $\fcno$, $\Ys$, the helium core mass and the hydrogen envelope mass of SN\,1987A. Initial mass ratios of $q \lesssim 0.2$ are indeed consistent with the assumption of a binary merger \cite{Schuermann2024}.

\subsection*{B: Why $\gamma$ Col is not a mass donor}\label{app_gamma}

It has been suggested that this star was stripped in a binary, revealing layers close to the initial convective core boundary that are in CN-equilibrium and only partially processed by the full CNO-cycle (fig.\,3 of \cite{Irrgang2022}). To assess this possibility, we examine mass donors in \Fig{fig_CNO2}. Our mass donors do not show particularly high N/C for a moderate N/O, rather, they lie close to the ``CNO-eq. + dilution''-line. On the most enriched end, they reproduce the surface abundances of Wolf-Rayet stars (e.g., a WN star WR\,40 \cite{Herald2001} and the WN/C star WR\,121-16 \cite{Zhang2020}). This is because binary mass stripping does not halt at the boundary of the initial convective core, but digs deeper into the inner region, until the surface hydrogen abundance is significantly reduced. At this stage, the stripped star reveals CNO-equilibrium matter. So, while the layers in the stripped single star model of \cite{Irrgang2022} which fit to the surface abundances of $\gamma$ Col will also be revealed during a binary mass transfer process, they will appear at the surface of the donor star for a short amount of time while mass transfer is ongoing. This   is certainly not the case in $\gamma$ Col.

The stripped star scenario proposed for $\gamma$ Columbae \cite{Irrgang2022} faces three more caveats. First, the star would not stay at $\gamma$ Columbae's position in the Hertzsprung-Russell diagram for long; the proposed model is currently undergoing mass stripping at that position and is out of thermal equilibrium. Second, it should have a nearby bright companion, which is not observed. In contrast, our mass gainer scenario offers a long-lived match to both the observed position in the Hertzsprung-Russell diagram and the surface abundances, while naturally explaining the lack of a companion. Third, it does not show peculiar surface gravity with respect to normal B-type stars, which is expected for stripped stars.

\subsection*{C: Mass gainer birth locations in the Hertzsprung-Russell diagram}\label{app_HRD}

\begin{figure}
	\centering
	\includegraphics[width=0.8\linewidth]{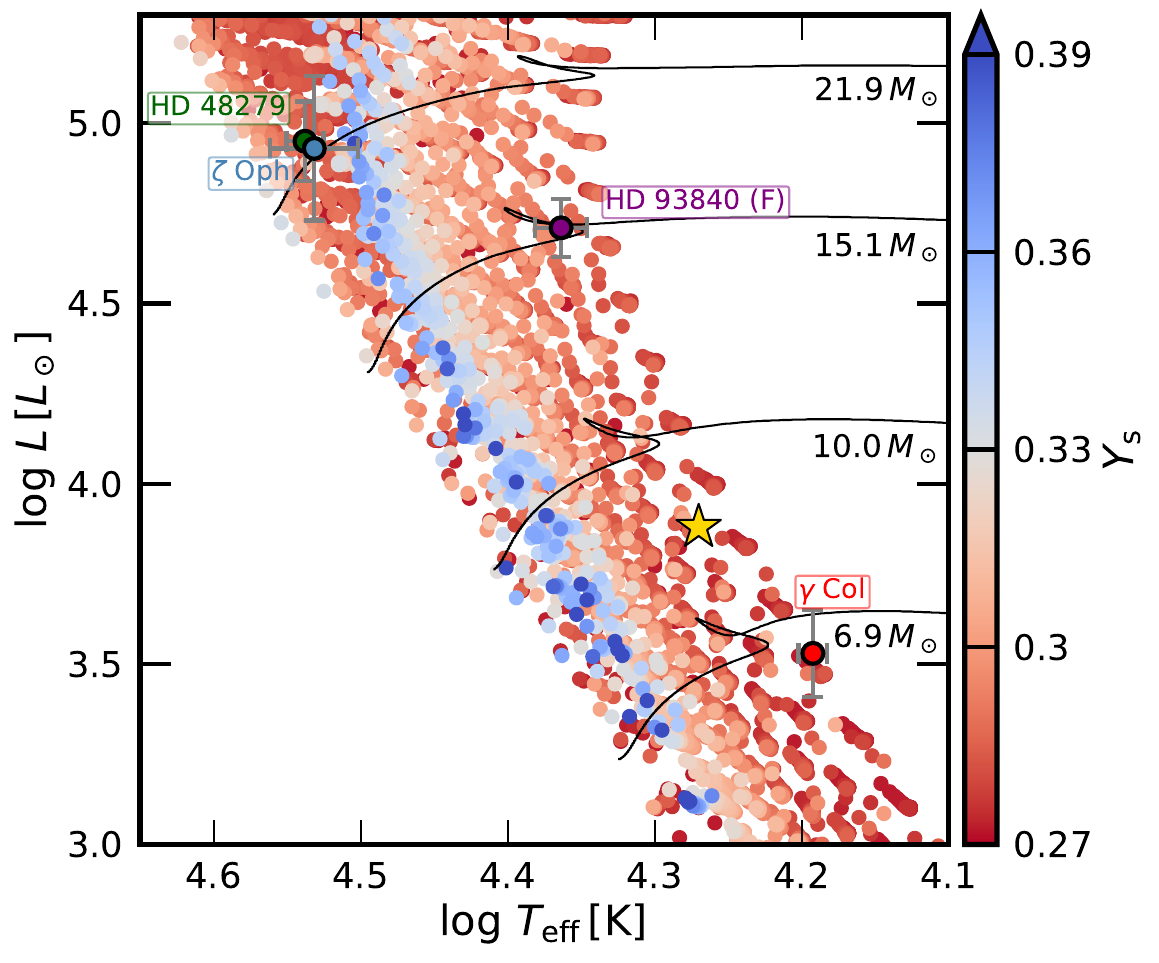}
	\caption{{\bf Mass gainers after thermal relaxation during core hydrogen burning}. Each data point represents one mass gainer model after at least one thermal timescale has passed since the completion of mass transfer. The color-coding represents surface helium mass fraction. Evolutionary tracks for non-rotating single star models are presented with their initial mass indicated. The crosses represent HD\,48279 \cite{Martins2015}, $\zeta$ Ophiuchi \cite{Villamariz2005}, $\gamma$ Columbae \cite{Irrgang2022}, and HD\,93840 \cite{Wessmayer2024}, and the star symbol represents our ``Mock star'' (see text).
    }
	\label{fig_HRD}
\end{figure}

Supplementary Figure~\ref{fig_HRD} shows our mass gainer models at the moment when the star has thermally relaxed after mass accretion. From this point onward, they evolve on the main sequence band and further into the supergiant region while keeping their surface helium abundances constant (see fig.\,3 of \cite{Jin2025}). In other words, helium-enriched, core hydrogen burning stars will span the entire main sequence band. The four observed stars investigated in this work are overplotted, and they lie on the main sequence. Beyond core hydrogen exhaustion, a slight increase in helium abundance may occur during the red supergiant stage.

\subsection*{\rev{D: Core size and CN-processing}}\label{app_ov}

\begin{figure}
	\centering
	\includegraphics[width=0.9\linewidth]{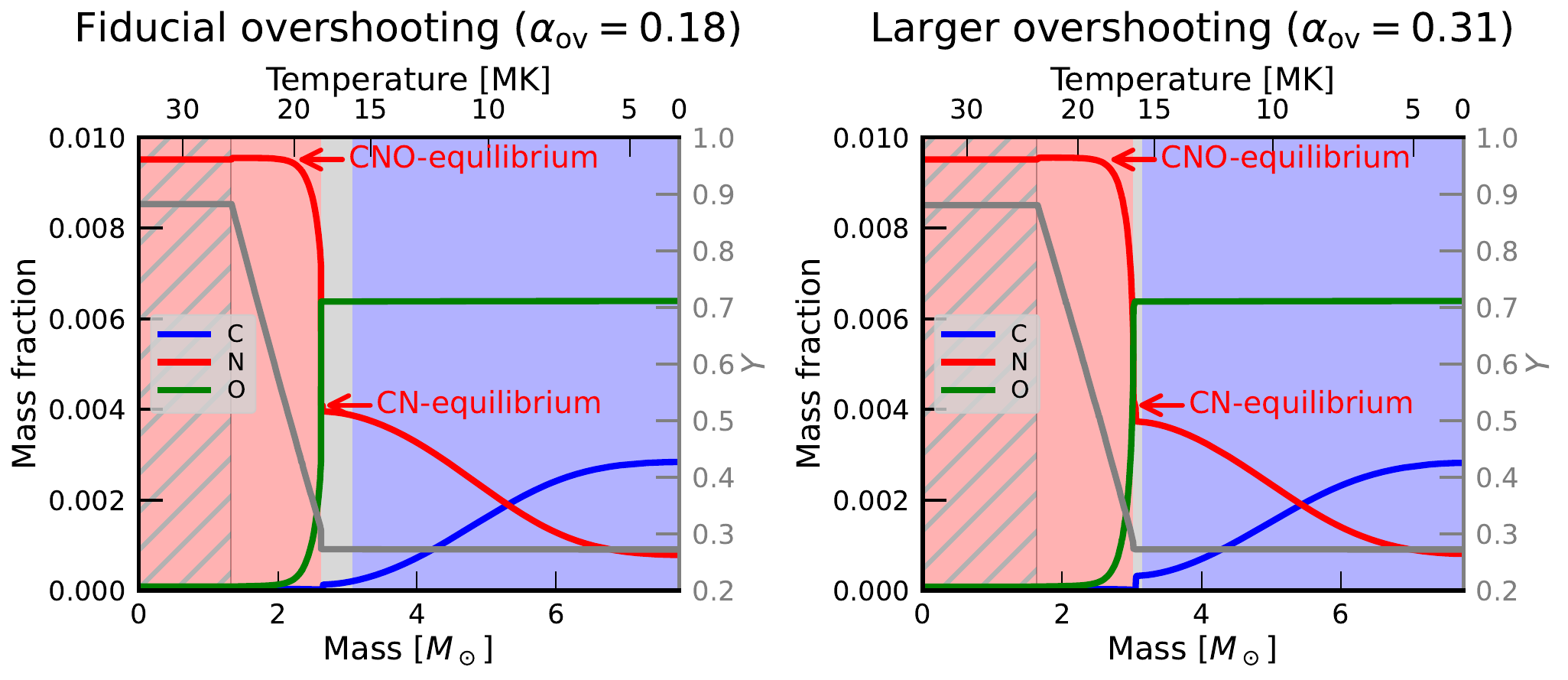}
	\caption{\rev{{\bf Internal profiles of helium and CNO elements for different overshooting parameters.} Mass fractions of CNO (left $y$-axis) and helium (right $y$-axis) of single star models of the same mass (7.8 $\mso$) computed with different overshooting parameters. Shown are the profiles when the central hydrogen mass fraction is 0.1. The red shaded region is the former initial convective region which contains matter in CNO-equilibrium, the hatched region is the current convective region, the grey region is where the CN-processing timescale is less than the main sequence lifetime, and the blue shaded region is above it. Red arrows indicate CNO- and CN-equilibrium nitrogen abundances. Left: From a model with fiducial overshooting (this work). Right: From a model with about twice larger overshooting \cite{Brott2011}. 
}
    }
	\label{fig_ov}
\end{figure}

\rev{Supplementary Figure~\ref{fig_ov} shows shows the effect of different core overshooting on the CN-processing region at the bottom of the envelope in single star models. It shows that the extent of the region at the bottom of the envelope (on top of the H/He-gradient region) that reaches near complete CN-equilibrium (gray shading) is smaller for larger overshooting. However, incomplete CN-cycling which produces significant nitrogen enrichment extends over $\sim2\mso$, such that the overall effect on nitrogen enhancement is almost independent of overshooting. Rejuvenation will also not affect the majority of these layers. Thus, regardless of whether the CN-equilibrium is reached at the bottom of the envelope, accretion-induced mixing will still raise the surface N/C ratio and put the mass gainers above the ``CNO-eq. + dilution''-line. 
}

\end{document}